\documentclass[a4paper,fleqn,usenatbib]{mnras_orig}

\usepackage{paralist,color,pdflscape}

\usepackage[T1]{fontenc}
\usepackage{ae,aecompl}
\usepackage[normalem]{ulem} % strikethrough option. This package can be deleted later.

\usepackage{graphicx}	% Including figure files
\usepackage{amsmath}	% Advanced maths commands
\usepackage{amssymb}	% Extra maths symbols

\title[X-ray census of the interacting binaries in NGC\,188]{A {\em Chandra} X-ray census of the interacting binaries in old open clusters---NGC\,188}
\author[S. Vats et al.]{Smriti Vats$^1$,\thanks{E-mail: s.vats@uva.nl; mvandenberg@cfa.harvard.edu}
Maureen van den Berg$^{2}$ and Rudy Wijnands$^{1}$\\
$^{1}$Anton Pannekoek Institute for Astronomy, University of Amsterdam, Science Park 904, 1098 XH Amsterdam, The Netherlands\\
$^{2}$Harvard-Smithsonian Center for Astrophysics, 60 Garden Street, Cambridge, MA 02138, USA}
\date{Accepted 2018 August 17. Received 2018 August 17; in original form 2018 June 21.}

\pubyear{2018}

\begin{document}
\label{firstpage}
\pagerange{\pageref{firstpage}--\pageref{lastpage}}
\maketitle

\begin{abstract}
We present a new X-ray study of NGC\,188, one of the oldest open
clusters known in our Galaxy (7~Gyr). Our observation with the {\em
Chandra X-ray Observatory} is aimed at uncovering the population of
close interacting binaries in NGC\,188. We detect 84 sources down to a
luminosity of $L_X$ $\approx$ 4$\times$10$^{29}$ erg~s$^{-1}$
(0.3--7~keV), of which 73 are within the half-mass radius $r_h$. Of
the 60 sources inside $r_h$ with more than 5 counts, we estimate that
$\sim$38 are background sources. We detected 55 new sources, and
confirmed 29 sources previously detected by {\em ROSAT} and/or {\em
XMM-Newton}. A total of 13 sources detected are cluster members, and 7
of these are new detections: four active binaries, two blue straggler
stars (BSSs), and, surprisingly, an apparently single cluster member
on the main sequence (CX\,33/WOCS\,5639). One of the BSSs detected
(CX\,84/WOCS\,5379) is intriguing as its X-ray luminosity cannot be
explained by its currently understood configuration as a
BSS/white-dwarf binary in an eccentric orbit of $\sim$120 days. Its
X-ray detection, combined with reports of short-period optical
variability, suggests the presence of a close binary, which would make
this BSS system a hierarchical multiple. We also classify one source
as a new cataclysmic-variable candidate; it is identified with a known
short-period optical variable, whose membership to NGC\,188 is
unknown. We have compared the X-ray emissivity of NGC\,188 with those
of other old Galactic open clusters. Our findings confirm the earlier
result that old open clusters have higher X-ray emissivities than
other old stellar populations.
\end{abstract}


\section{Introduction} \label{ch3_intro}

X-ray production in single late-type stars is powered by a dynamo
mechanism in their convective zones, which means that the faster the
star rotates, the higher is its X-ray luminosity. According to the
Skumanich law \citep{skumanich72}, the rotational velocities of single
low-mass stars are proportional to the reciprocal of the square root
of their age: as stars get older, they tend to slow down due to
magnetic braking \citep{Pallavicini:1989p1112}, and correspondingly
their X-ray emission decreases. Our Sun is one such old star
($\sim$4.5 Gyr) that has an X-ray luminosity of about $10^{26 - 27}$
erg s$^{-1}$ (0.1--2.4 keV; \citealt{Peres:2000p1107}). Such X-ray
luminosities are nearly undetectable beyond distances of about a
kiloparsec with the current generation of X-ray telescopes, not even
with the sensitivity of the \textit{Chandra X-ray
  Observatory}. However, X-ray observations of old (age $\gtrsim$ 1
Gyr) open clusters have detected a rich population of X-ray sources
that are associated with these clusters. Follow-up studies showed many
of these sources to be close binaries of late-type stars that have
been spun up by tidal interaction \citep{Belloni:1993p1075,
  Gondoin:2005p1051, Giardino:2008p1048,
  Gosnell:2012p685,vandenBerg:2004p1040, vandenBerg:2013p442,
  vatsvdb2017}. Such tidally interacting binaries, also known as
active binaries (ABs), can have either two detached stars comprising
the binary, or can have a contact or semi-detached configuration like
W\,UMa and Algol binaries, respectively. X-ray sources in old clusters
can also be accretion-powered, as in the case of cataclysmic variables
(CVs), where a white-dwarf primary accretes matter from a late-type
main-sequence donor. These CVs are typically found to the blue of the
main sequence in the colour-magnitude diagram (CMD) due to the light
from the accretion disk or stream, and possibly due to the
contribution from the white dwarf itself.  There are also some exotic
X-ray sources found in old open clusters, like blue straggler stars
(BSSs) and sub-subgiants (SSGs), however the origin of the X-ray
emission from these sources, as well as the evolutionary status of
these stars themselves, is not well understood
(e.g.~\citealt{geller+17}). Old open clusters are very good
laboratories to study binaries, such as ABs, CVs, SSGs and BSSs, as it
is possible to obtain cluster membership information---and therefore
age and distance---for a large number of sources with much less effort
than for X-ray sources in the Galactic field. Also, the stellar
densities of open clusters lie between those of dense globular
clusters (GCs; $\gtrsim 10^4 M_{\odot}$ pc$^{-3}$) and the solar
neighbourhood ($\sim0.1 M_{\odot}$ pc$^{-3}$). Hence, studying old
open clusters aids us in understanding the role of stellar density in
stellar and binary evolution. \citet{verbunt2000} found that the old
open cluster M\,67 has a higher X-ray emissivity than most
GCs. \citet{geea2015} demonstrated the elevated X-ray emissivity of
two open clusters, viz. NGC\,6791 and M\,67, with respect to old
stellar populations other than GCs, like dwarf galaxies and the local
solar neighbourhood. In \citet{vatsvdb2017} we were further able to
improve the statistics when we found that the old open cluster
Collinder 261 (Cr\,261; age $\sim7$ Gyr) is also over-luminous in
X-rays compared to a few dense GCs.

To expand our understanding of open-cluster X-ray sources and the
evolution of binaries in different environments, we are undertaking a
survey with {\em Chandra} of old open clusters with ages between 3.5
Gyr and 10 Gyr. The observations are designed to reach a limiting
luminosity of at least $L_{X} \approx 10^{30}$ erg s$^{-1}$ (0.3 --
7~keV). In the current paper, we focus on the binary population of
NGC\,188, which at an estimated age of $\sim$7 Gyr
\citep{sarajedini99}, is one of the oldest open clusters in the
Galaxy. It lies at a distance of 1650$\pm$50 pc, has half-mass and
core radii of $r_h=8\farcm3\pm0\farcm6$ and $r_c=4\farcm4\pm0\farcm1$
respectively, and a reddening of $E(B-V) = 0.083$
\citep{Chumak:2010p1841}. NGC\,188 was first observed in X-rays using
the $ROSAT$ Position Sensitive Proportional Counter (PSPC) with a flux
detection limit of $\sim$10$^{-14}$ erg cm$^{-2}$ s$^{-1}$ in the
energy range 0.1--2.4~keV \citep{Belloni:1998p1070}, equivalent to a
luminosity limit of $\sim4\times10^{30}$ erg s$^{-1}$ for the quoted
distance. The cluster was re-observed with {\em XMM-Newton} as a
performance verification object with a luminosity threshold of
$\sim$10$^{30}$ erg s$^{-1}$ (0.5--2.0~keV;
\citealt{Gondoin:2005p1051}), but the cluster was not centred at the
aimpoint during the observation, leading to asymmetric coverage of the
cluster. Recent studies by the WIYN Open Cluster Survey (WOCS) have
led to significant progress in our knowledge of NGC\,188. Membership
of the cluster was established using proper-motion studies performed
by \citet{platais2003} and stellar radial-velocity measurements
performed by \citet{geller2008}. We use both of these studies for
determining cluster membership of the likely and candidate optical
counterparts to the {\em Chandra} sources detected in the study we
present here.

NGC\,188 hosts one of the best-studied populations of BSSs
(e.g.\,\citealt{geller2011}, \citealt{gosnell2014,gosnell2015}). BSSs
are stars that are bluer and brighter than the main-sequence turnoff
point of a coeval population. They were first discovered in the
globular cluster M\,3 by \citet{sandage53}, however, our understanding
of their formation is still quite poor. There are currently three
suggested scenarios for how BSSs are formed -- mass transfer in a
binary system, merger of two (or more) stars due to a direct
collision, and merger of the inner close binary induced by a tertiary
companion in a hierarchical triple (see \citet{Davies2015book} for a
review). Uncovering blue-straggler formation scenarios therefore
contributes to our understanding of stellar encounters in clusters.
Not all BSSs are expected to be sources of X-ray emission. However, if
X-rays are observed from a particular BSS this could give away
valuable clues to its current configuration (e.g.\,it may indicate the
presence of a close binary in the system) and thereby constrain its
past evolution.

In Section \ref{ch3_obs_ana} we describe the observations and
analysis. In Section \ref{ch3_results} we explain how we performed the
source classification and we present our results. In Section
\ref{ch3_discussion} we discuss our findings, and give details about
two particularly interesting sources, including a BSS (WOCS\,5379)
whose X-ray emission is not well understood. Section \ref{ch3_summary}
is a summary of this work.

\begin{figure}
\includegraphics[clip=,width=1.0\columnwidth]{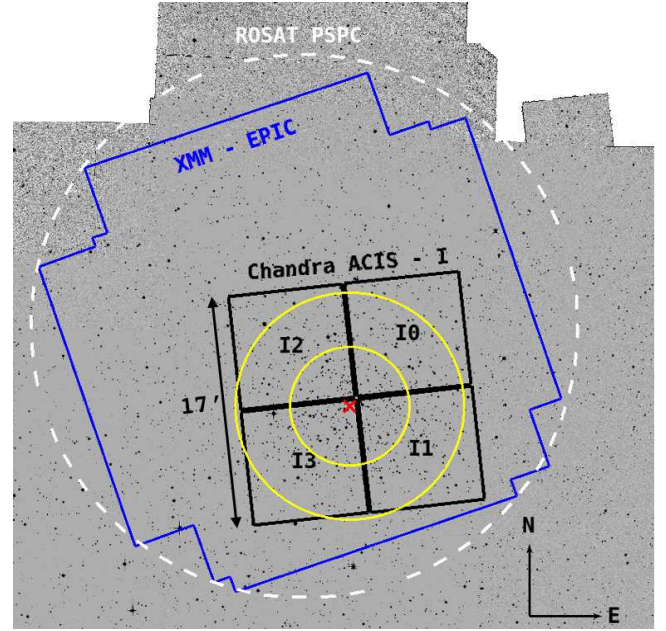}
\caption{$V$-band image of NGC\,188 and its surroundings from
  \citet{stetson2004}. The white dashed circle shows the central
  20\arcmin\,of the $ROSAT$ PSPC observation of NGC\,188. The blue
  composite region shows the area covered by the {\em XMM-Newton} EPIC-PN and
  EPIC-MOS detectors. The black squares show the area covered by the
  {\em Chandra} ACIS-I chips. The red cross marks the centre of the cluster,
  and the outer and inner yellow circles mark the half-mass radius,
  $r_h$, and the core radius, $r_c$, respectively.}
 \label{ch3_fov}
\end{figure}

\section{OBSERVATIONS AND ANALYSIS} \label{ch3_obs_ana}

\subsection{X-ray Observations}  

NGC\,188 was observed with the Advanced CCD Imaging Spectrometer
\citep[ACIS;][]{Garmire:2003p984} on board {\em Chandra} starting on 2013 July
6 10:25 UTC and the total exposure time of the observation was 24.7 ks
(ObsID 14939). The observation was made in very-faint, timed exposure
mode, with a single frame exposure time of 3.1 s. The CCDs used were
I\,0, I\,1, I\,2 and I\,3 from the ACIS-I array, and S\,3 from the
ACIS-S array. The centre of the cluster ($\alpha_{2000} =
0^{h}47^{m}12.5^{s}$, $\delta_{2000} = +85^{\circ}14'\,49''$;
\cite{Chumak:2010p1841}) was placed close to the {\em Chandra} aimpoint on the
ACIS I\,3 chip, so that most of the cluster area within the half-mass
radius $r_h$ was imaged on the ACIS-I chips (see Figure \ref{ch3_fov} for
the chip layout).

We reduced the level-1 event file created by the standard processing
pipeline of the {\em Chandra} X-ray Center using the CIAO 4.5 routines and the
CALDB 4.5.5.1 calibration files. We use the \verb chandra_repro  reprocessing
script to obtain the level-2 event file. A background light curve for the
energy range 0.3--7 keV was created with the \verb dmextract  routine in
CIAO using source-free areas of the ACIS-I chips, and was analysed with the
\verb lc_sigma_clip  routine. We did not observe any background flares with
 more than 3$\sigma$ variations from the average background count
 rate. Therefore, we used events detected during the entire
 observation for our further analysis.

\subsection{X-ray Source Detection and Source Characterisation}

X-ray source detection and characterisation was performed in a similar
manner as was done for the cluster Cr\,261 by \citet{vatsvdb2017}. We
summarise the method here. X-ray analysis was performed only on data
obtained from chips I\,0, I\,1, I\,2 and I\,3, within the energy range
of 0.3--7~keV (broad energy band). This energy range was further
divided into soft (0.3--2~keV) and hard (2--7~keV) energy bands for
our analysis. We did not use data from chip S\,3, as S\,3 lies far
from the aimpoint, leading to large positional errors on sources
detected on this chip. These large errors make it difficult to
securely identify optical counterparts, and thus to classify the
sources.

The CIAO source detection routine \verb wavdetect  was run for eight
wavelet scales ranging from 1.0 to 11.3 pixels, each increasing by a
factor of $\sqrt{2}$. Exposure maps were computed for an energy value
of 1.5~keV. The \verb wavdetect  detection threshold ({\tt sigthresh})
was set at $ 10^{-7} $, for which the total number of spurious
detections expected is 3.35 for all four ACIS chips and eight wavelet
scales combined. For the three energy bands, we detected 72 distinct
X-ray sources with more than 2 counts (0.3--7 keV). To check if we had
missed any real sources, we ran \verb wavdetect  again for a more
lenient detection threshold of $10^{-6}$, and found 84 distinct X-ray
sources. The number of spurious detections expected for this threshold
for all wavelet scales, over all four chips is 33.5. One of the twelve
extra sources found for the detection threshold of $ 10^{-6} $ matches
with the known variable V\,04 \citep{Hoffmeister64}, which is a
short-period W\,UMa variable and a plausible X-ray source. We flagged
these extra twelve sources, because the majority are likely to be
spurious, but have kept them in the master X-ray source list.

For cross-correlating the optical and X-ray catalogues, we calculate
the positional uncertainties on the X-ray sources by using the 95\%
confidence radius on their {\tt wavdetect} positions, $P_{err}$
\citep{Hong:2005p959}. We determined the net source counts using ACIS
Extract (AE) \citep[version 2013mar6]{Broos:2010p966} as {\tt
  wavdetect} is not optimised for this task. All events in the energy
range of 0.3--7 keV were extracted. The extraction regions enclose
$\sim$90\% of the PSF at 1.5 keV. For sources with five counts or more
that spend more than 90\% of the total exposure time on the ACIS-I
detector (56 sources in total), AE performs variability
characterisation based on a Kolmogorov-Smirnov (K-S) test on the event
arrival times. Based on the K-S variability test, there are 52 sources
with no evidence for variability $(0.05 < P_{KS})$; two sources that
showed possible variability $(0.005 < P_{KS} < 0.05$; CX\,66, CX\,75)
and two that were definitely variable $(P_{KS} < 0.005$; CX\,1,
CX\,11), where $P_{KS}$ is the probability of having a constant count
rate. Neither CX\,1 nor CX\,11 is a member of NGC\,188. The optical
counterpart of CX\,1 is an optical photometric variable---V\,08
\citep{kaluznyshara1987}---for which a binary origin was considered in
\citet{mazurkaluzny1990}. We classify CX\,11 as a likely background
active galactic nucleus (AGN; Section \ref{ch3_sec_nonmem}).

Only two sources in our catalogue (CX\,1 and CX\,2) have more than 100
net counts (0.3--7 keV) with the brightest source CX\,1 having 256 net
counts. For most of the other sources, the spectral shape of the X-ray
emission is poorly constrained because of the low number of counts
detected. We calculated the unabsorbed X-ray fluxes, $F_{X,u}$,
using the \verb Sherpa  package in CIAO. We assumed a 2~keV MeKaL model
(\textit{xsmekal}) and a neutral hydrogen column density,
$N_{H}=5\times10^{20}$ cm$^{-2}$ (obtained using the adopted $E(B-V)$ for
NGC\,188, and the conversion between $A_V$ and $N_H$ given in
\citealt{Predehl:1995p1114}), using the {\em xstbabs} model. A MeKaL
model is appropriate for ABs as it describes the emission from a hot,
diffuse gas or optically thin plasma typical for stellar coronae
\citep{Gudel:2004p12}. Since we do not know the underlying spectrum of
many of our sources, we explored the effect of using different
spectral models on the values of $F_{X,u}$. We compared the values of
the unabsorbed flux obtained using the 2~keV MeKaL model with those
obtained using a 1~keV MeKaL model (\textit{xsmekal}), a 10~keV
thermal bremsstrahlung model (\textit{xsbrems}) and a power law model
(\textit{xspowerlaw}) having a photon index $\Gamma=1.4$. We used the
same absorption model and value in all cases ($xstbabs$). The
unabsorbed flux values were, on average, about 14\% smaller, 26\%
larger and 33\% larger than the value for the 2~keV MeKaL model,
respectively.

We characterise the spectral properties of our X-ray sources using the
method of quantile analysis \citep{Hong:2004p940}. In this method, the
median energy, $E_{50}$, and 25\% and 75\% quartile energies of the
event energy distribution ($E_{25}$ and $E_{75}$, respectively) of a
source are used to determine its hardness and spectral shape. For
sources with few photon counts, conventional hardness ratios, which
use pre-defined hard and soft energy bands, may not give meaningful
results if all those events lie only in either one of the
bands. Details of the source properties are presented in
Table\,\ref{ch3_tab1}, and the quantile diagrams are shown in
Figure\,\ref{ch3_quantiles}.

\subsection{Optical Source Catalogue and X-ray and Optical Cross-matching} 

We created an optical master catalogue for NGC\,188 by combining the
catalogues provided in \citet{platais2003} and
\citet{stetson2004}. The positional errors in \citet{platais2003} are
different for each source and are between 1.9 mas and about 100 mas,
both in right ascension and declination. The \citet{stetson2004}
catalogue is a combined catalogue from 11 older data sets, leading to
different positional errors for different sources. However, they are
expected to be less that 0\farcs1 in both right ascension and
declination. The photometry provided in both catalogues was calibrated
onto the standard Johnson $BV$ system. The $B$ and $V$ magnitudes, and
the optical coordinates used in our study are obtained from
\citet{platais2003} except for the sources that are only present in
\citet{stetson2004}. The range in $V$ magnitude in the
\citet{platais2003} catalogue is $9.0<V<22.0$ and in $B$ magnitude it
is $9.0<B<23.1$.

Despite the fact that {\em Chandra} sources have good absolute astrometric
accuracy\footnote{The 95\% confidence radius on absolute positions
  from {\em Chandra} ACIS-I is 0$\farcs$9 -- 1\arcsec\,within a distance of
  3\arcmin\,from the aimpoint; see
  http://cxc.harvard.edu/cal/ASPECT/celmon.}, there is still a chance
that there is a systematic offset in {\em Chandra} source positions compared to
our optical positions that have been calibrated to the International
Celestial Reference System (ICRS), i.e.~a boresight, which complicates
the search for optical counterparts if not corrected for. To calculate
the boresight, we first identified 66 known short-period ($P <$ 5
days) variables and one FK\,Com source among the sources in our
optical master catalogue that lie within the {\em Chandra} field of view of
NGC\,188. Many surveys for optical variables in NGC\,188 have been
performed, and we have compiled a list of variables from the following
references: \cite{kaluznyshara1987}, \cite{kafhon2003},
\cite{zhang2004}, and \cite{mochejska2008}. Short-period binaries are
expected to be tidally locked, increasing their X-ray
activity. FK\,Com sources are rapidly rotating single stars believed
to be the outcome of a binary merger, and this rapid rotation leads to
increased X-ray emission. Hence, short-period variables and FK\,Com
stars are likely X-ray emitters and have lower chances of being
spurious matches to X-ray sources (see also Section 2.4). We then
cross-matched the ICRS-calibrated optical positions of these variables
with our {\em Chandra} catalogue using a 95\% match radius, which is the
combination of the error in the optical positions given in the
\citeauthor{platais2003} catalogue scaled to a 95\% confidence radius,
and the random error on the X-ray positions ($P_{err}$). Thus, we
found 16 counterparts to the X-ray sources which we then used to
calculate the boresight. Once an initial boresight was calculated, the
cross-matching is repeated and the match radii are augmented with the
boresight error. This step was repeated until the net boresight
converged. The boresight was found to be 0\farcs12$\pm$0\farcs11 in
right ascension and --0\farcs14$\pm$0\farcs13 in declination, which is
consistent with zero. These values were then added to the old X-ray
positions. The method for calculating and correcting for the boresight
can be found in more detail in Section 3.3.1 of
\citet{vandenBerg:2013p442}.

We matched the boresight-corrected X-ray source list with the optical
source list using 95\% error radii. Of the 84 unique X-ray sources
detected within the field of view, 35 were matched with a single
optical counterpart while one source (CX\,76) was matched with two
optical counterparts. A complete list of candidate counterparts for
these 36 X-ray sources can be found in Table \ref{ch3_tab2}. Candidate
optical counterparts are presented in the CMD of
Figure\,\ref{ch3_hrd}.

On manual inspection of the $V$ image provided by \cite{stetson2004},
we find that CX\,76 appears to have four more faint optical
counterparts, including two that look extended. Of the 48 X-ray
sources without any counterparts in the optical catalogue, 12 sources
appear to have optical counterparts that were too faint to be reported
by \cite{platais2003} or \cite{stetson2004}. However, we do not report
these sources in Table \ref{ch3_tab2} as we do not have photometric
information for these faint optical sources.

\begin{figure*}
\centerline{
%\\
\includegraphics[width=7cm,angle=90]{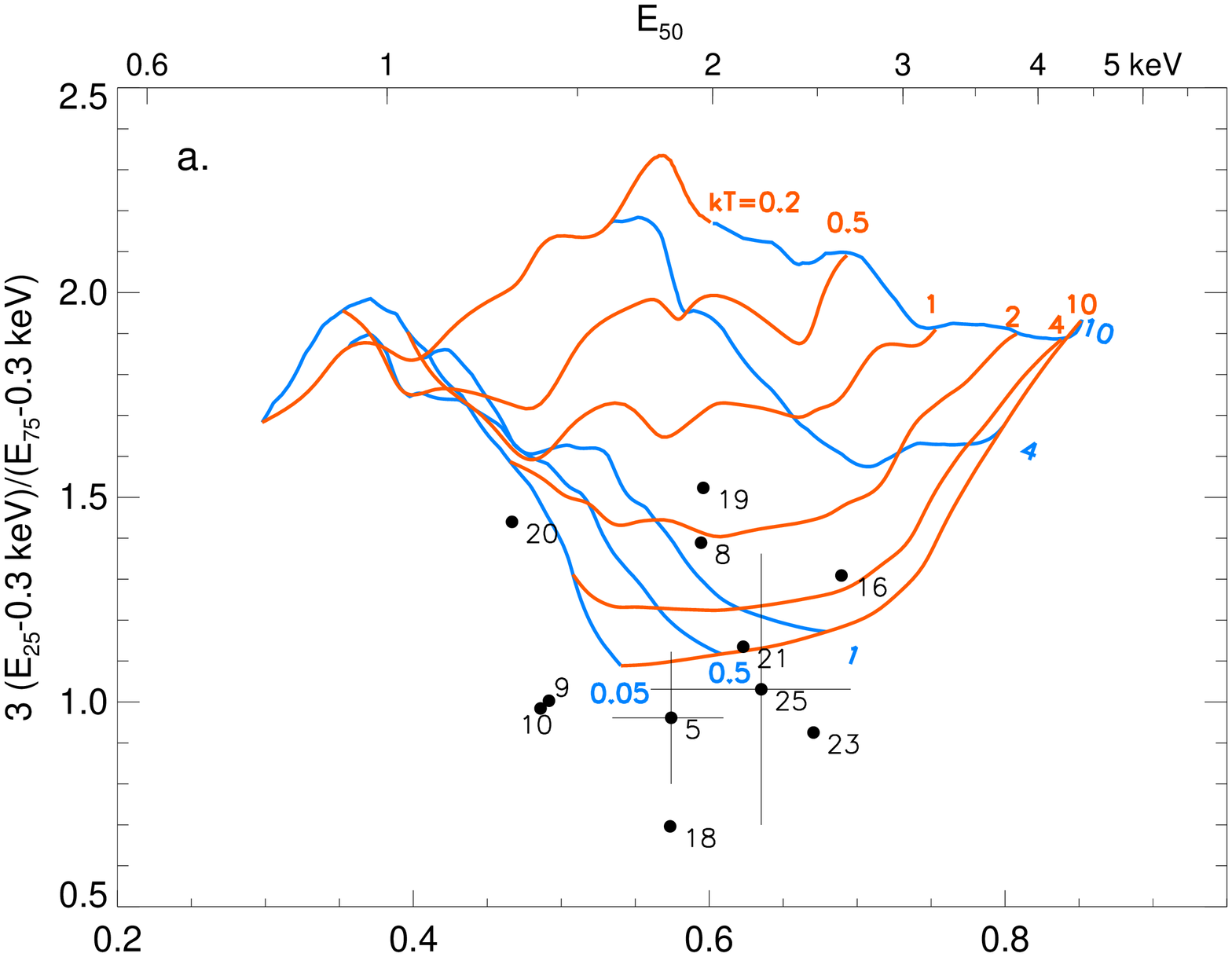}
\hspace{-1.cm}
\includegraphics[width=7cm,angle=90]{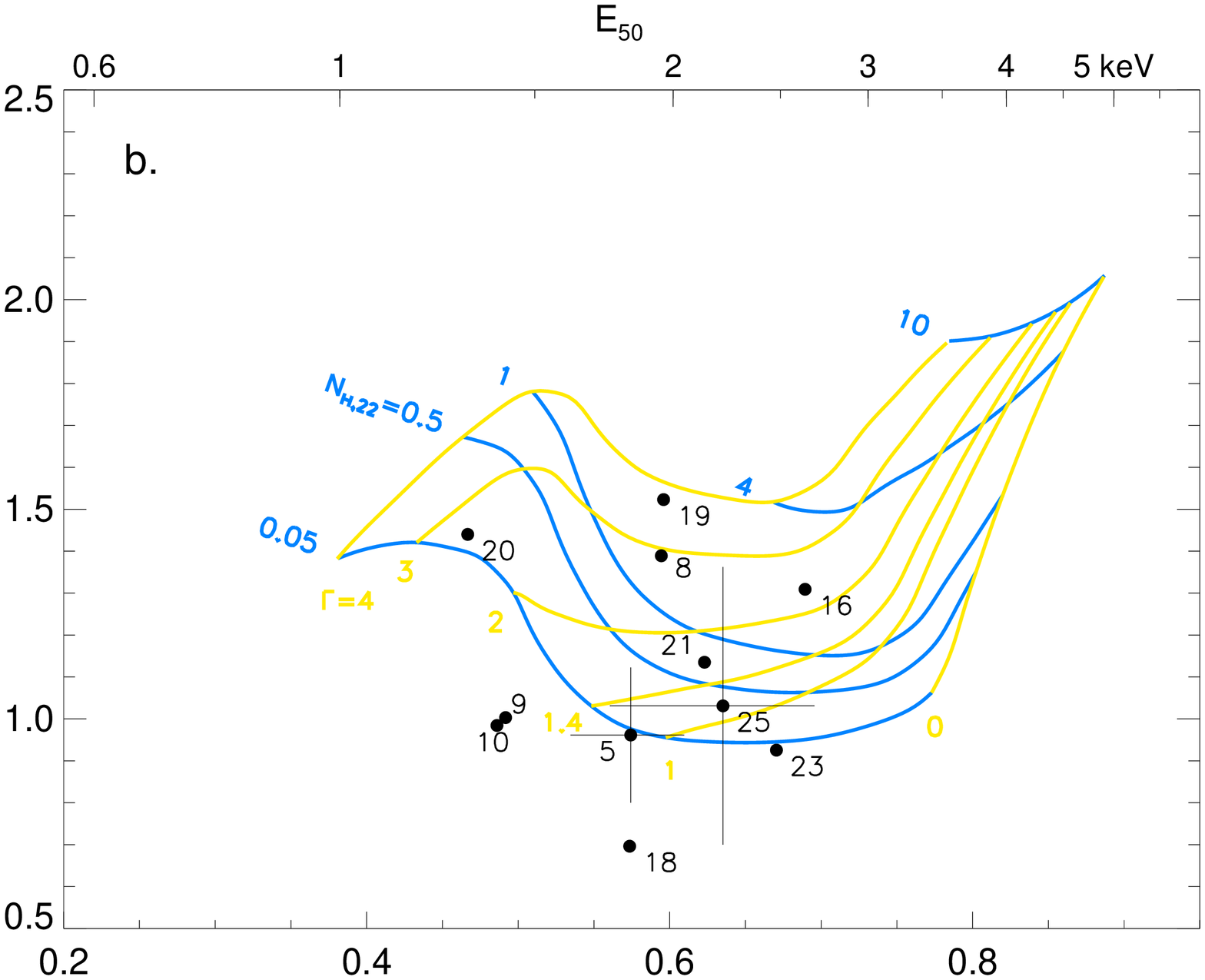}
}
\end{figure*}
\begin{figure*}
\centerline{
%\\
\includegraphics[width=7cm,angle=90]{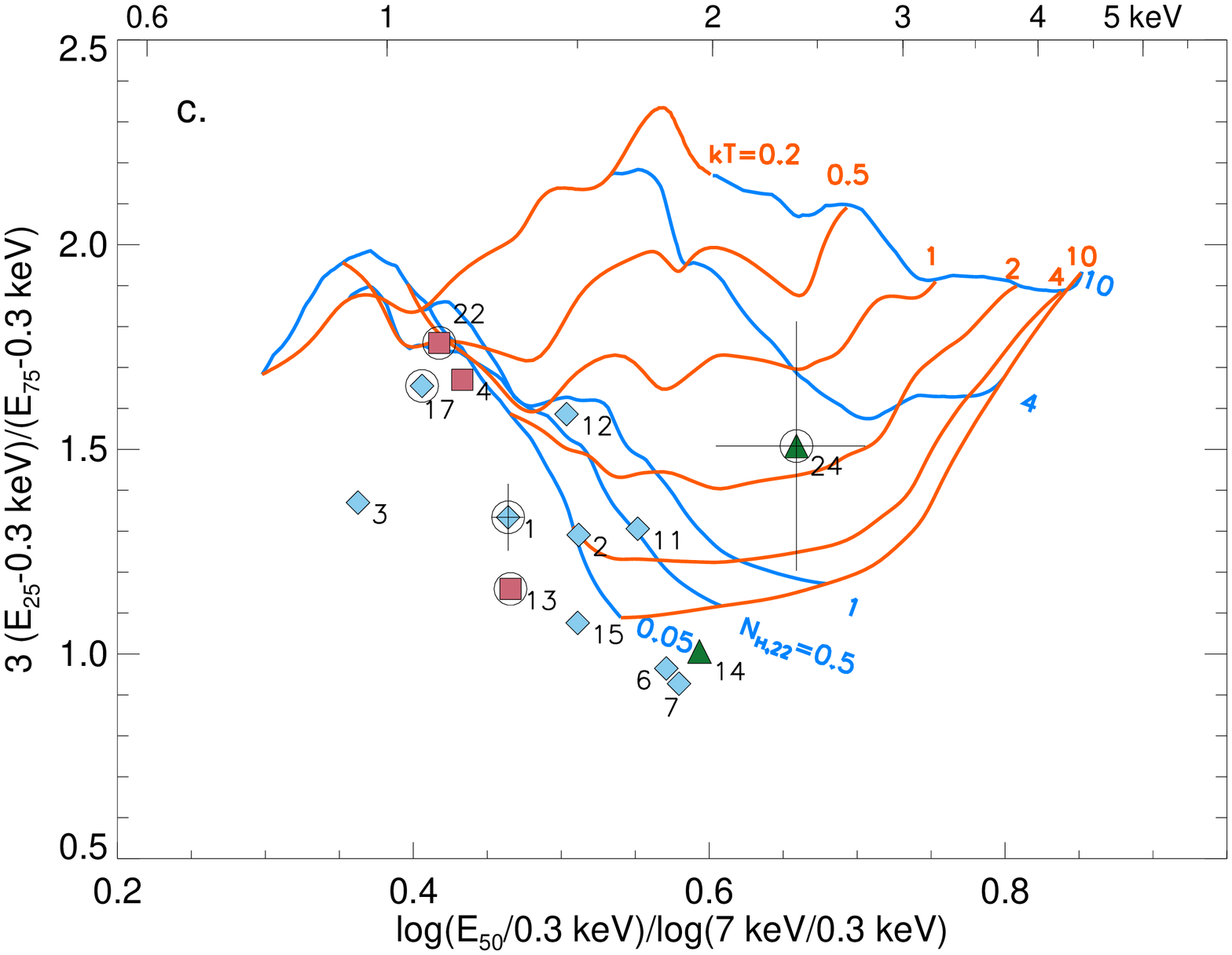}
\hspace{-1.cm}
\includegraphics[width=7cm,angle=90]{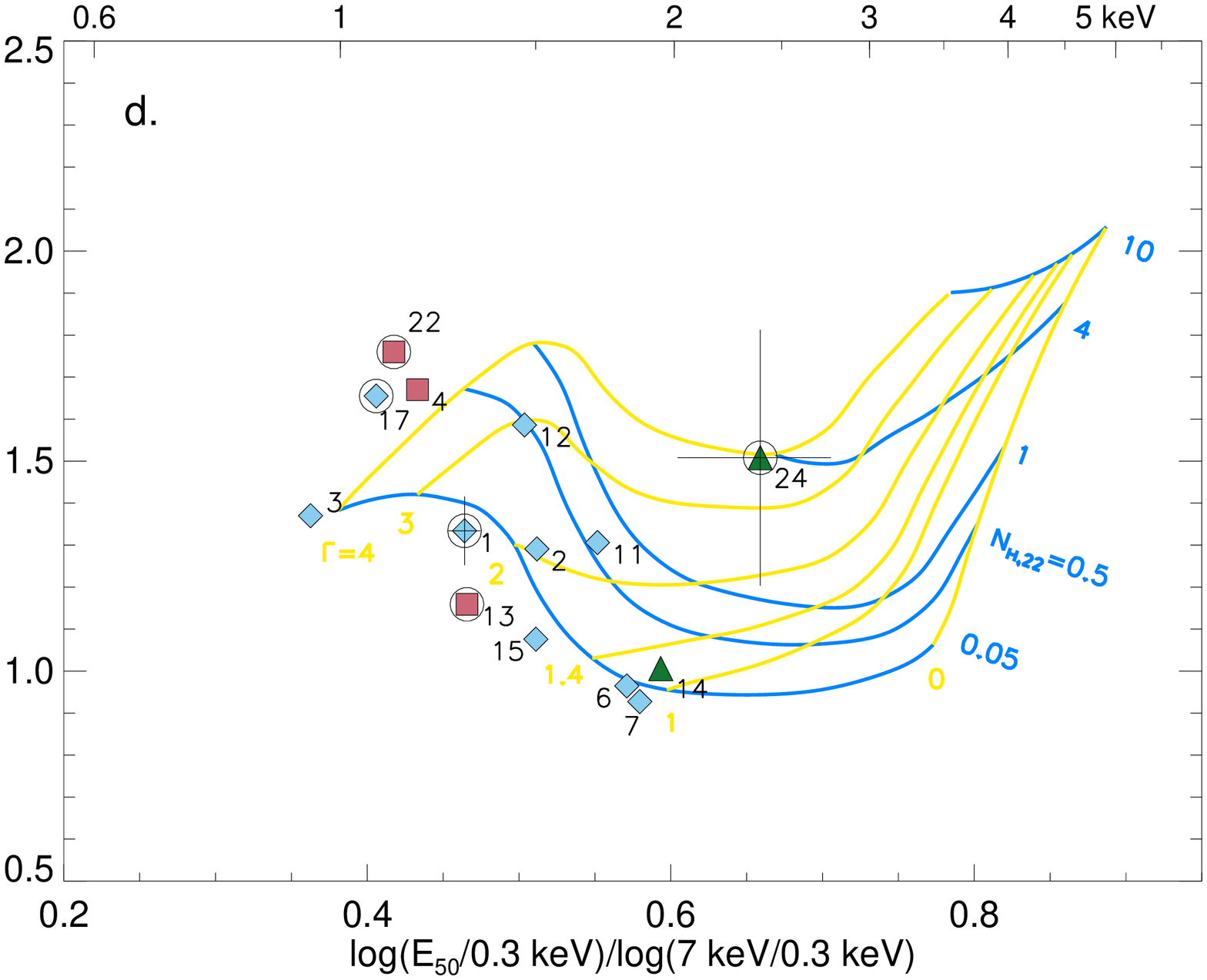}
}
\caption{Quantile diagrams with model grids representing a MeKaL
  plasma (left) and a power-law spectrum (right). The top panels show
  sources without candidate optical counterparts and the bottom panels
  show sources with candidate optical counterparts (see Table
  \,\ref{ch3_tab2}). For a certain choice of model, the plasma
  temperature (\textit{kT}) or photon index ($\Gamma$), and the column
  density ($N_{H}$) can be estimated from the location of a source
  inside the grid. The median energy \textit{E$_{50}$} can be read off
  from the top x-axis. Blue lines in all the four quantile diagrams
  represent different values of $N_{H}$ normalised in units of
  $10^{22}$ ($N_{H,22}$), where $N_{H,22}\approx0.05$ cm$^{-2}$ is the
  cluster value. Here we show sources with 20 net counts (0.3--7~keV)
  or more; error bars are shown only for the sources with the highest
  and lowest number of counts. In the bottom panels, red squares are
  members, green triangles have uncertain membership and blue diamonds
  are likely non-members. Sources with known periodic photometrically
  variable counterparts are circled with a larger black open circle.}
 \label{ch3_quantiles}
\end{figure*}

\begin{figure*}
  \includegraphics[clip=,width=2.0\columnwidth]{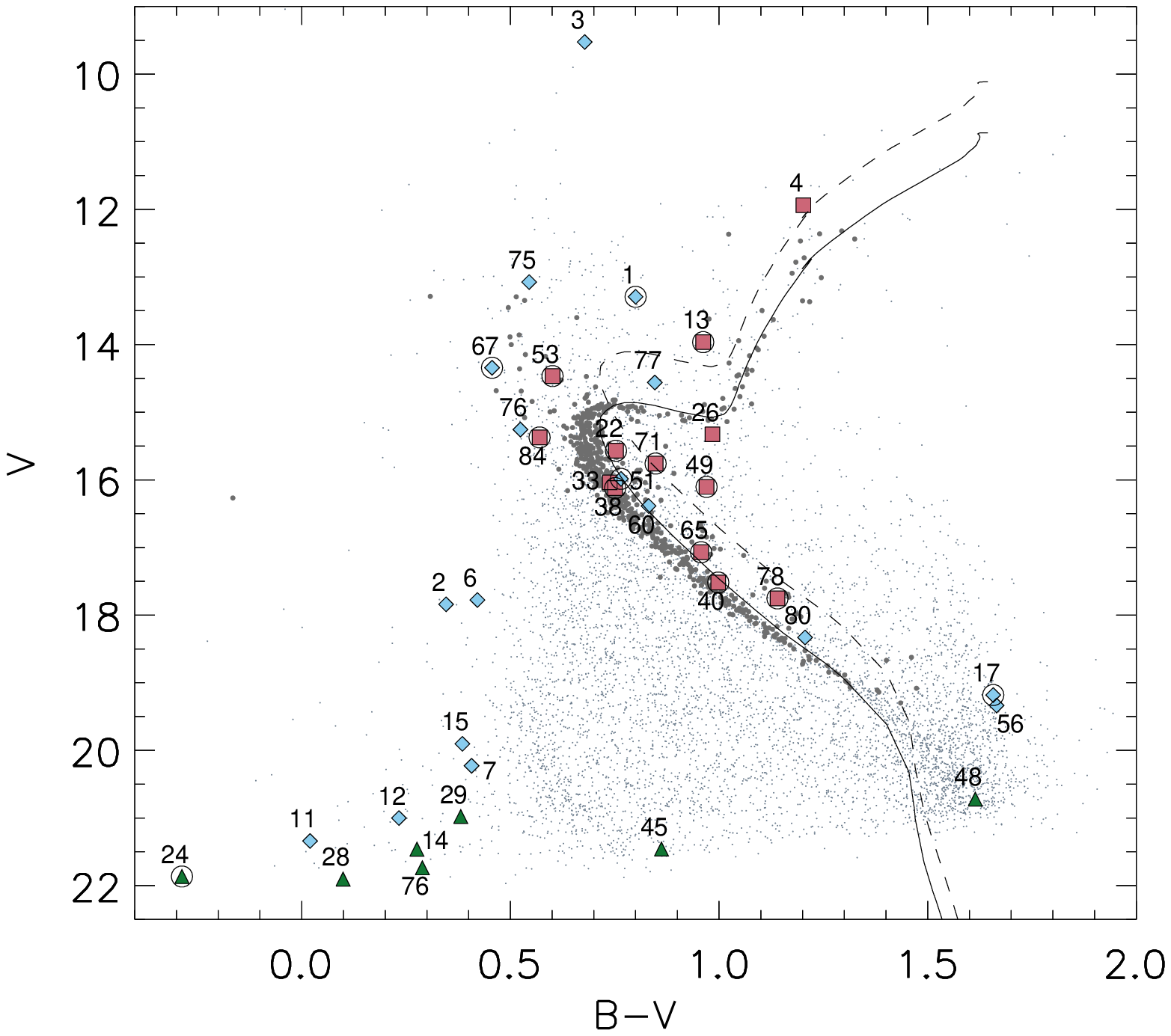}
\caption{Colour-magnitude diagram of NGC\,188. Stars with confirmed
  membership based on proper-motion (MP$_{PM}>90\%$) are marked with a
  bold grey dot while the remaining are marked with normal grey
  dots. Filled colour symbols are candidate optical counterparts to
  {\em Chandra} sources. Among them red squares are cluster members,
  green triangles have uncertain membership and blue diamonds are
  likely non-members (MP$_{PM}$ and/or MP$_{RV}$ is 0\%). Furthermore,
  candidate counterparts that are known binary counterparts are
  circled with a larger black open circle. The solid line represents
  the isochrone
  \citep{Bressan:2012p1120,Chen:2014p1422,Tang:2014p1207,Chen:2015p452}
  for the reddening associated with the cluster ($E(B-V)=0.083$). The
  dashed line represents the isochrone shifted upwards by $-0.75$
  magnitudes to indicate the limit for photometric main sequence
  binaries.}
\label{ch3_hrd}
\end{figure*}

\subsection{False Positives Test and Background Galaxies} \label{ch3_bkg}

To estimate the number of spurious matches between our X-ray and
optical sources, we calculated the stellar density of optical sources
in the cluster field using our optical source catalogue. Inside the
core radius $r_c$ as measured by \citet{Chumak:2010p1841}, the average
density is 0.0023 arcsec$^{-2}$, while in the annulus between $r_c$
and the half-mass radius $r_h$ the average density is 0.001
arcsec$^{-2}$.

Multiplying the stellar densities with the area covered by the 95\%
match radii around the X-ray sources within these regions we expect
0.24 spurious matches among the thirteen matches that we find in the
central region, and 0.98 spurious matches among the seventeen matches
in the outer region. We perform the same calculations in order to
estimate the number of spurious matches between X-ray sources and
photometric variables. The stellar density of variables inside $r_c$
is 0.00012 arcsec$^{-2}$, while between $r_c$ and $r_h$ the average
density is $5.6\times10^{-5}$ arcsec$^{-2}$. We expect 0.013 spurious
matches among the nine matches that we find in the central region,
and 0.040 spurious matches among the six matches that we find in the
outer region. Therefore, all matches with short-period variables are
likely real.

To estimate the number of background galaxies $N_B$ among the sources
in our X-ray catalogue, we used the relation for the cumulative number
density of high--galactic-latitude X-ray sources above a given flux
limit (Eq.~5 in \citealt{Kim:2007p933}). We adopted the $\log N - \log
S$ relation for the $B$ band (0.3--8 keV), which is closest to our
broad band (0.3--7~keV). To convert counts to fluxes we assumed a
power-law spectrum with $\Gamma=1.4$. As NGC\,188 lies far above the
plane of the Galaxy, the integrated Galactic column density in the
direction of NGC\,188 is the same as the Galactic column density
towards the cluster itself \citep{Drimmel:2003p1555}; therefore we
assume $N_H=N_{H,NGC\,188}$. Closer to the cluster centre, the density
of cluster stars is higher and mass segregation makes the radial
distribution of binaries more concentrated
\citep{sarajedini99,kafhon2003,geller2008}. So we calculated $N_B$ for
$r<r_c$, where most X-ray sources that are truly associated with
NGC\,188 are expected to be. For a detection limit of 5 counts, we
expect $N_B\approx9.3$$\pm$3.1 versus 21 sources actually
detected. For a detection limit of 10 counts, it is expected that
$\sim$4.9$\pm$2.2 of the eight sources detected within $r_c$ are
extragalactic. In the region $r_c < r \leq r_h$, 21.9$\pm$4.7 of the
39 sources detected above 5 counts, or 11.3$\pm$3.4 of the 29 sources
detected above 10 counts are expected to be extragalactic.

\subsection{Comparison with other X-ray observations}

NGC\,188 was previously observed with $ROSAT$
\citep{Belloni:1998p1070} with a detection limit of
$\sim4\times10^{30}$ erg s$^{-1}$ (0.1--2.4 keV), and with {\em
  XMM-Newton} \citep{Gondoin:2005p1051} with a limit of
$\sim1\times10^{30}$ erg s$^{-1}$ (0.5--2.0 keV). Both observations
were less sensitive compared to our {\em Chandra} observation in which
the faintest detected cluster member has $L_X\approx 4\times 10^{29}$
erg s$^{-1}$ (0.3--7 keV)\footnote{All luminosities are normalized to
  our adopted cluster distance of 1650 pc.}. We looked for X-ray
counterparts to our {\em Chandra} sources in these two catalogues. Out
of the 34 $ROSAT$ sources, nine are within the field of view (FOV) of
our {\em Chandra} observation and one is partially within the FOV. We
find that eight sources from the \citeauthor{Belloni:1998p1070}
catalogue (sources labelled with `X') match with nine of our {\em
  Chandra} sources (sources labelled with `CX') within a 95\% match
radius (X\,19/CX\,2, X\,20/CX\,5 X\,30/CX\,3, X\,29/CX\,4,
X\,21/CX\,21+CX\,22, X\,25/CX\,25, X\,17/CX\,39, X\,32/CX\,56) and one
more source matches within a $\sim$3.6-$\sigma$ match radius (CX\,1
matches with X\,26 within 6\farcs8; see Table\,\ref{ch3_tab1}). The
95\% match radius is a combination of the 95\% error radius on the
{\em Chandra} source position and the 90\% error radius on the $ROSAT$
source position, scaled to 95\%. However, one source (X\,27) in the
\citeauthor{Belloni:1998p1070} catalogue, that lies within the field
of view of our {\em Chandra} observation, has no counterparts in our
X-ray catalogue. This is unexpected as the detection limit of the {\em
  ROSAT} observation is an order of magnitude higher than that of our
{\em Chandra} observation.  It is plausible that X\,27 is a variable
source in, or projected onto, the cluster.  \citet{Gondoin:2005p1051}
finds 58 sources in NGC\,188 of which 31 are within the {\em Chandra}
FOV. In the case of the \citeauthor{Gondoin:2005p1051} catalogue
(sources labelled with 'GX'), errors on source position are not
provided, because of which finding cross-matches is not
straightforward. However, after recent refinements in the {\em
  XMM-Newton} data reduction algorithms, a catalogue of all X-ray
sources detected in {\em XMM-Newton} observations has been compiled by
the {\em XMM-Newton} Survey Science Centre. To find cross-matches
between {\em XMM-Newton} detections and {\em Chandra} detections, we
made use of the sixth data release of the 3XMM {\em XMM-Newton}
serendipitous source catalogue (3XMM-DR6) which contains source
detections from all {\em XMM-Newton} EPIC observations made between
February 3, 2000 and June 4, 2015 \citep{rosen2016}. We find 91 3XMM
sources in our {\em Chandra} FOV (compared to the 31 reported by
\citealt{Gondoin:2005p1051}; both source lists for this field are
based on the same observations), of which 41 get matched with 40 of
the {\em Chandra} sources within a 95\% matching radius (six more 3XMM
sources lie within a 3$\sigma$ radius of another six of the {\em
  Chandra} sources). Of these 41 matched sources, only nineteen were
previously reported by \citet{Gondoin:2005p1051}; see
Table~\ref{ch3_tab3}.

\begin{figure}
\includegraphics[clip=,width=1.0\columnwidth]{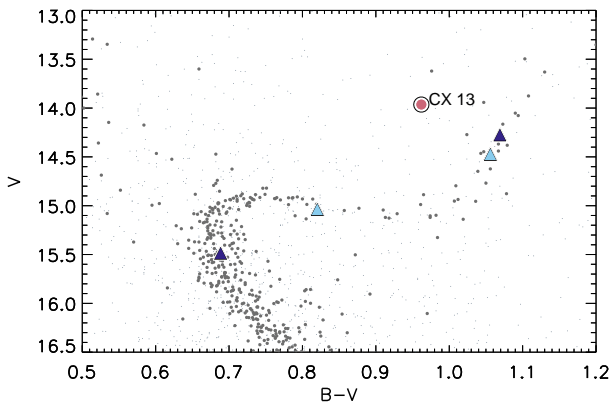}
\caption{Photometric decomposition of the optical counterpart of
  CX\,13 into its plausible components. The two light blue triangles
  represent one pair of plausible binary components and the two dark
  blue triangles represent another pair.}
\label{ch3_decomp}
\end{figure}

\section{Results} \label{ch3_results}

For classifying our X-ray sources, we use three diagnostics, details
of which can be found in \citet{vatsvdb2017}. In short, we first look
at the hardness of the X-ray spectrum, which can be inferred from the
position of the source in the quantile diagram
(Figure\,\ref{ch3_quantiles}). For sources with fewer than 20 counts
we infer the hardness from the median energy, \textit{E$_{50}$}, as
the quantile diagram no longer provides reliable information about the
underlying spectral shape of sources with few net X-ray counts. As the
integrated Galactic $N_H$ is the same as the $N_H$ towards the
cluster, Galactic X-ray sources without any intrinsic absorption
should not have an $N_H$ larger than the cluster value. Also,
coronally active stars and binaries usually have temperatures of about
3--4 keV or less. In the MeKaL grid in Figure\,\ref{ch3_quantiles},
one can see that a combination of these temperature and $N_H$
constraints gives $E_{50} \lesssim 1.5$ keV. For CVs and AGNs $E_{50}$
may be higher due to their harder intrinsic spectra.

Next, we looked at the ratio of the unabsorbed X-ray flux to optical
flux $(F_X/F_V)_u$ for sources with candidate optical counterparts, or
the lower limit on this ratio for sources without optical
counterparts. We adopted a 2 keV MeKaL model to calculate X-ray
fluxes, assuming $N_H=N_{H,NGC\,188}$ to correct for absorption, and
using $V_0=V-N_{H,NGC\,188} / (1.79 \times 10^{21})$
\citep{Predehl:1995p1114} for the unabsorbed $V$ magnitude. The
calculated flux ratio is overestimated if the adopted $N_H$ is lower
than the actual $N_H$ or underestimated if the adopted $N_H$ is higher
than the actual $N_H$. This situation may arise as we do not know the
$N_H$ for most sources beforehand. The \textit{$\log (F_X/F_V)_u$}
ratio also helps us distinguish between coronal and accretion-powered
sources. Typically, active stars and binaries have
\textit{$\log(F_X/F_V)_u \lesssim -1$}, while accretion-powered
sources have \textit{$\log(F_X/F_V)_u \gtrsim -1$}
\citep{Stocke:1991p305}.

Finally, for sources with candidate optical counterparts we also look
at the position of these optical sources in the CMD. In most cases,
this works reasonably well to separate active stars and binaries from
AGNs (which often have blue optical colours and lie far off the
stellar main sequence in the CMD) and CVs (which typically are blue,
too).

\subsection{Cluster Members}

To classify sources as cluster members or non-members, we considered
the membership probabilities of their candidate optical counterparts
from both proper-motion studies \citep[MP$_{PM}$;][]{platais2003} and
radial-velocity studies \citep[MP$_{RV}$;][]{geller2008}.  All optical
sources that matched with our {\em Chandra} sources and for which membership
information is available, fall in two well-separated membership
categories. Thirteen X-ray sources have optical counterparts with
MP$_{PM}$ and/or MP$_{RV}$ larger than 90\%; we consider these as
probable cluster members. For sixteen X-ray sources MP$_{PM}$ and/or
MP$_{RV}$ is 0\%; these sources are briefly discussed in Section
\ref{ch3_sec_nonmem}. In almost all cases, MP$_{PM}$ and MP$_{RV}$ are
consistent; the only exception is CX\,51 (Section
\,\ref{ch3_sec_nonmem}).

Of the thirteen X-ray sources that are likely members, six had
previously been detected by $ROSAT$ or {\em XMM-Newton} or both, and
seven are new detections (CX\,33, CX\,40, CX\,53, CX\,65, CX\,71,
CX\,78, CX\,84). All new detections lie within $r_h$, with four lying
within $r_c$. In the following subsections, we first discuss the ABs
(Section \,\ref{ch3_sec_ab}), and then the anomalous stars like BSSs
(Section \,\ref{ch3_bsssg}) and an uncertain classification (Section
\,\ref{ch3_unc}).

\subsubsection{Active binaries} \label{ch3_sec_ab}

For selecting possible ABs in NGC\,188, we picked sources with optical
counterparts that lie along the cluster main sequence or sub-giant
branch in the CMD. We allowed for the possible contribution to the
optical light from binary companions. For a binary with two companions
of the same mass, the maximum decrease in $V$ magnitude is 0.75 and
the resulting binary sequence is indicated by the dotted isochrone in
Figure\,\ref{ch3_hrd}. We also considered their X-ray properties, and
photometric or radial-velocity periods. Seven {{\em Chandra}} sources were thus
selected as likely ABs, of which the brightest three were detected
before (CX\,13/GX\,18, CX\,22/X\,21, CX\,38/GX\,20). All have $E_{50}$
values $\lesssim1.5$ keV, \textit{$\log F_X/F_V$} $\lesssim -1.8$, and
are matched with short-period ($\lesssim11$ days) photometric or
radial-velocity variables. We have classified these sources as ABs in
NGC\,188, and classify them as `AB's in column 14 of
Table\,\ref{ch3_tab2}.

We add a few notes on individual sources. CX\,71 is matched to the
W\,UMa-type contact binary V\,04 \citep{Hoffmeister64,mochejska2008};
we confirm its cluster membership using the period-colour relationship
for W\,UMa's \citep{Rucinski:1994p862}. The counterpart to the AB
CX\,13 (V\,11) lies in a region of the CMD that is $\sim0.15$ mag
bluer than the red giant branch. As noted by several authors, this
location can be explained by a combination of a relatively unevolved
red giant and a less-evolved member. We illustrate this in
Figure\,\ref{ch3_decomp}, where we photometrically decompose the
optical source in to its plausible binary components, i.e.\,we
investigate the fluxes of different types of stars that can be
combined to give a total flux equal to the flux of CX\,13. We further
discuss this source in Section \ref{ch3_sec_cx13}. The counterpart to
CX\,22 (V\,21) was classified by \citet{zhang2004} as a W\,UMa binary
based on its light curve; with a photometric period of
$P_{ph}\approx1.17$ days, they note it is the longest-period W\,UMa in
NGC\,188. \citet{mochejska2008}, however, find a period of
$P_{ph}\approx1.38$ days (assuming one maximum/minimum per cycle,
instead of two as would be the case for W\,UMa's) and classify it as a
BY\,Dra system. To be a W\,UMa, its period would have to be $\sim2.77$
days, which is decisively too large for a W\,UMa star. We thus adopt
the period by \citeauthor{mochejska2008}, also because the light curve
shown in \citet{zhang2004} does not show the typical W\,UMa-type
shape.

\subsubsection{Blue stragglers, sub-subgiants and FK\,Com stars}\label{ch3_bsssg}

The two BSSs in NGC\,188 detected by {\em Chandra} are both new X-ray
detections. The BSS counterpart of CX\,53 (WOCS\,5078 or V\,34) was
first observed as a variable by \citet{mochejska2008} with a period of
$\sim$4.5 days.  \citet{geller2008} found it to be a double-lined
spectroscopic binary with a period of $\sim$4.8 days. The orbital
period is short enough for strong tidal coupling and enhanced X-ray
emission, but interestingly, the orbit is eccentric ($e\approx0.12$)
pointing at a recent dynamical encounter or a third companion (see
discussion in \citealt{geller2009}). CX\,84 (WOCS\,5379 or WV\,3) is a
BSS with a white-dwarf companion \citep{gosnell2014}. The difference
between the reported radial-velocity ($P_{sp}\approx120$ days;
\citealt{geller2009}) and photometric ($P_{ph}\approx0.18$ or
$P_{ph}\approx0.36$ days; \citealt{kafhon2003}) periods is
intriguing. We discuss CX\,84 in more detail in Section
\ref{ch3_cx84}.

CX\,26 (GX\,28) and CX\,49 (GX\,45) are classified as SSGs. This
nomenclature arises from their location on the CMD, wherein they lie
below the subgiant branch and to the red of the main sequence. Several
scenarios have been proposed for how SSGs arrive at their current CMD
position (see \citet{geller+17} and the references within). CX\,49 is
the variable V\,05 discovered by \citet{kaluznyshara1987} with
$P_{ph}\approx0.59$ days. The lightcurve for V\,05 does not appear to
have equal maxima and minima, suggesting it may not be a W\,UMa binary
as suggested in \citet{mochejska2008}, but a semi-detached binary.

CX\,4 (X\,29) is a red giant and the brightest cluster member detected
in X-rays; it was classified as an FK\,Com variable by
\citet{harris1985} and a single cluster member by \citet{geller2009}.

\subsubsection{CX\,33: uncertain classification} \label{ch3_unc}

The optical counterpart to the new source CX\,33 lies on the cluster
main sequence, has a median energy value of $E_{50}\approx1.5$ keV,
and an X-ray--to--optical flux ratio
\textit{$\log(F_{X}/F_{V})_{u}$}$=-2.14$. These properties seem to
suggest an AB nature. However, according to \cite{geller2008}, the
optical counterpart to this source (WOCS\,5639) is a single cluster
member. They note that stars classified as single may be binaries that
cannot be detected as such because they have low-amplitude
radial-velocity variations and/or long orbital periods. A long-period
binary with main-sequence components, however, is not expected to be
an X-ray source as there is little interaction between the two stars
in such a system. Hence, we place CX\,33 under the category of
`uncertain classification'.

\subsection{Uncertain Membership}

Sources that do not have a membership probability based on either
proper motion or radial velocity, are classified as uncertain members
of the cluster. We found seven such sources of which two were detected
previously by {\em XMM-Newton} (CX\,24/GX\,57; CX\,28/GX\,31) and five are
newly detected sources (CX\,14; CX\,29; CX\,45; CX\,48; CX\,76).

Six of the seven uncertain members are proposed as CV or AGN
candidates as they lie to the blue of the cluster main sequence and
have \textit{$\log(F_{X}/F_{V})_{u}$}$ > -0.46$. Of these six, CX\,24
has the highest possibility of being a CV, as it is also matched with
the short-period variable WV\,2 \citep{kafhon2003} with a photometric
period of 0.15049 days.  For these sources, either confirmation of
cluster membership (or non-membership) or optical spectroscopy can
help establish their nature.

For the one remaining source, classification is uncertain. CX\,48 is
found at the lower end of the cluster main sequence and has a high
value of the median energy, $E_{50}=4.3\pm0.7$ keV. Also, its
X-ray--to--optical flux ratio, \textit{$\log(F_{X}/F_{V})_{u}$}$
\approx -0.7$, lies in the range for accreting binaries, or active
M-stars. If CX\,48 is at the distance of NGC\,188, its X-ray
luminosity is $1.4\times10^{30}$ erg s$^{-1}$. This source could be an
AB that underwent a flare during the {\em Chandra}
observation. AcisExtract did not detect X-ray variability from C\,X48,
but with only 10 counts the sensitivity for detecting variability is
low.

\subsection{Cluster non-members} \label{ch3_sec_nonmem}

Sixteen sources with candidate counterparts that have MP$_{PM}=0\%$ or
MP$_{RV}=0\%$ are considered to be non-members. These sources are most
likely a mix of foreground stars (those with bright counterparts and
soft X-ray spectra) and AGNs (with blue/faint counterparts, e.g.~the
X-ray variable CX\,11). In case of CX\,51, where MP$_{PM}$ is 98\% but
MP$_{RV}$ is 0\%, we consider the radial-velocity information to be
more constraining and hence, classify the source as a non-member. No
proper motion or radial velocity has been reported for the counterpart
to CX\,3, but at $V=9.5$ this star is too bright to be a cluster
member.

\subsection{Sources without candidate optical counterparts}

For 48 sources we do not find any candidate optical counterparts in
the \citeauthor{platais2003} or the \citeauthor{stetson2004}
catalogues. The detection limit ($V\approx22$) of the
\citeauthor{platais2003} catalogue allows us to place approximate
lower limits on their X-ray--to--optical flux ratios. These limits
range from \textit{$\log(F_X/F_V)_{u,lim}$}$ \approx -0.15$ for the
faintest (CX\,83) to \textit{$\log(F_X/F_V)_{u,lim}$}$ \approx 1.2$
for the brightest (CX\,5) unmatched source. The average $E_{50}$ for
the unmatched sources is 2.2$\pm$0.6 keV (versus 1.6$\pm$0.8 keV for
sources that do have candidate counterparts). These properties are
consistent with an AGN nature, an explanation that also looks
plausible when considering the number of unmatched sources. Among the
60 sources inside $r_h$ with $\geq$5 net counts, 30 do not have any
candidate optical counterparts. Given that we expect 31.2$\pm$5.6 AGNs
in this area with $\geq$5 counts (see Section \ref{ch3_bkg}), and that
we have identitifed $<$10 candidate AGNs among the sources {\em with}
candidate counterparts inside $r_h$, it is likely that the majority of
the sources without an optical match are AGNs.

\begin{figure}
  \includegraphics[width=8.5cm]{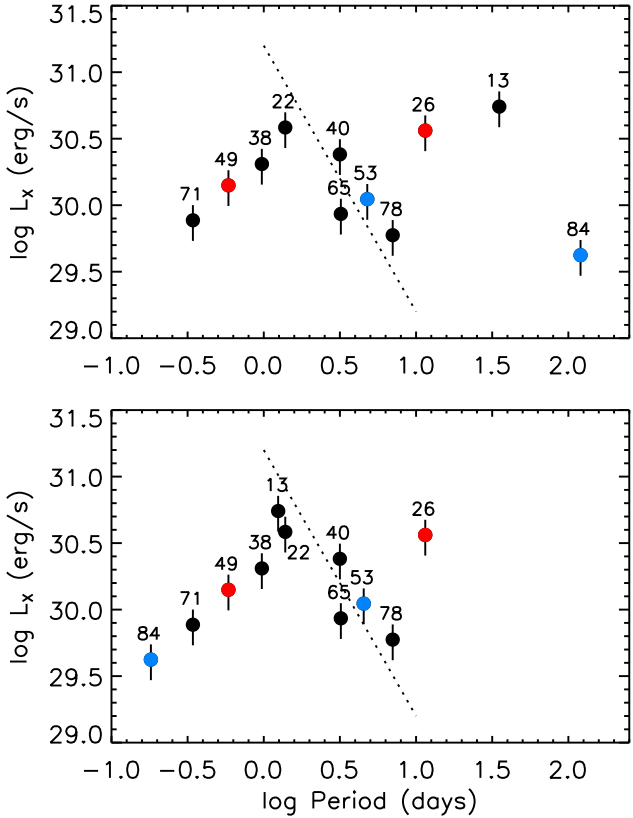}
  \caption{X-ray luminosity ($L_X$) versus rotation period ($P_{rot}$)
  plot, where either $P_{ph}$ or $P_{sp}$ is taken as a measure of
  $P_{rot}$.  In the top panel, $P_{rot}$ is based on radial
  velocities ($P_{sp}$) if available, and otherwise on variability
  studies. In the bottom panel, the periods are from variability
  studies ($P_{ph}$) if available, and otherwise from radial
  velocities. Blue stragglers are marked with blue circles,
  sub-subgiants with red circles. The dashed line shows the relation
  $L_X \propto P_{rot}^{-2}$ with an arbitrary vertical offset; note
  this is {\em not} a fit to the data. $L_X$ is based on fluxes
  computed for an assumed 2 keV MeKaL spectrum and the cluster $N_H$;
  errors on $L_X$ stem mostly from the unknown underlying spectrum,
  and are of the order of up to 20--30\%. Error bars of 30\% in $L_X$
  are plotted, while the formal errors on the periods are negligible.}
\label{ch3_pvslx}
\end{figure}

\section{Discussion}\label{ch3_discussion}

We conducted an X-ray study of the old open cluster NGC\,188 using
{\em Chandra}. Our study is several times more sensitive than
observations of the cluster done with {\em ROSAT}
\citep{Belloni:1998p1070} and {\em XMM-Newton}
\citep{Gondoin:2005p1051}. Unlike in these previous studies, the {\em Chandra}
pointing was centred on the cluster centre, providing a more symmetric
coverage of the cluster. Also, the positional accuracy of {\em
  Chandra} helps us in identifying candidate optical counterparts to
the X-ray sources. Of the 84 X-ray sources detected by {\em Chandra,}
35 unique X-ray sources have a single candidate optical counterpart,
while one X-ray source (CX\,76) has at least two candidate optical
counterparts. Of the remaining 48 X-ray sources, 12 appear to have
candidate optical counterparts in the $V$-band image by
\citet{stetson2004}, however, these faint sources are neither included
in the \citet{stetson2004} nor in the \citet{platais2003} catalogue.

As shown by \citet{Mathieu:2004p602}, late-type main-sequence binaries
in NGC\,188 show a transition around $P_{sp}=15.0$ days from circular
orbits (below this period) to eccentric orbits (for longer
periods). This demonstrates that at 7 Gyr (i.e.\,the age of NGC\,188)
tidal interaction has had sufficient time to circularise orbits up to
this cutoff period. Since tidal synchronisation of the stars' rotation
occurs before circularisation (e.g.\,\citealt{Zahn:1989p220}), we
expect late-type binaries with orbital periods at least up to 15 days
to have achieved tidal coupling ($P_{rot} = P_{sp}$) and have enhanced
levels of X-ray emission compared to single stars of similar age and
mass. Indeed, all the ABs discussed in Section\,\ref{ch3_sec_ab} have
periods shorter than 15 days, except CX\,13 for which the
radial-velocity period is longer (discussed below).

The X-ray activity of a late-type main-sequence star is related to its
rotation period \citep{pizzolato03}.  Generally, the X-ray luminosity
increases with decreasing rotation period and follows the relation
$L_X \propto P_{rot}^{-2}$.  However, for short periods a star reaches
a ``saturated'' regime, where $L_X$ no longer depends on $P_{rot}$,
but instead flattens to $L_X/L_{bol} \sim 10^{-3}$ with $L_{bol}$
being the bolometric luminosity of the star
\citep{randich98,Gudel:2004p12}.  Stars with the shortest periods
appear to emit X-rays at a level even lower than predicted by this
saturated luminosity relation. Such stars are said to be
``super-saturated'' \citep{prosser96} and this behaviour has been
ascribed to either a change in the dynamo action of the star or the
reduction in the area of the active regions (see Sect.~5.4 in
\citealt{Gudel:2004p12}). In Figure\,\ref{ch3_pvslx}, we plotted $L_X$
versus a measure of $P_{rot}$ for cluster members that were matched
with a {\em Chandra} source to see if we observe this behaviour in
NGC\,188. The periods used for the sources on the top panel were taken
from the radial-velocity orbital solutions from \cite{geller2009} if
available (CX\,13, CX\,26, CX\,53, CX\,84), and otherwise from
variability surveys \citep{mochejska2008,zhang2004}.  We can see that
the overall behaviour of $L_X$ as function of period is in line with
the trend described above, albeit with significant scatter. Moving
from $\sim$7 days to shorter periods, $L_X$ reaches higher values
until $\sim$1.3 days; this is similar to what we see for the ABs in
M\,67 \citep{vandenBerg:2004p1040}. For shorter periods, $L_X$ starts
diminishing again--this may correspond to the super-saturation
regime. Three sources are outliers: CX\,13, CX\,26 and CX\,84. For
CX\,13 and CX\,84 the period from radial velocities is different from
the reported photometric period. If we plot the latter instead (bottom
panel), these two sources are more in line with the trend
described. CX\,13 and CX\,84 are described in more detail in Sections
\ref{ch3_sec_cx13} and \ref{ch3_cx84}. The SSG CX\,26 is more X-ray
luminous for its period compared to the overall trend; the
spectroscopic period is the only measure for the rotation period
available.

There are nine cluster members within the {\em Chandra} FOV with
orbital periods from \citet{geller2009} below $\sim15$ days, that were
not detected in X-rays. Six of these sources could simply have X-ray
luminosities below our detection limit as their periods are relatively
long ($8$ days $< P_{sp} < 15$ days). In addition, WOCS\,5762 ($P_{sp}=6.5$
days) and WOCS\,5147 ($P_{sp}=6.7$ days) are found further away from
the aimpoint where the X-ray detection limit is higher. However, we
cannot explain the non-detection of WOCS\,5052. This spectroscopic
binary lies closer to the aimpoint; it has a 3.85 day spectroscopic
period and a near-circular orbit ($e=0.05\pm0.03$;
\citealt{geller2009}).

\subsection{CX\,13}  \label{ch3_sec_cx13}

The optical counterpart to CX\,13 (WOCS\,4705) is an eccentric ($e =
0.487\pm0.005$), double-lined spectroscopic binary with an orbital
period of $P_{sp}\approx35.2$ days \cite{geller2009}. Its orbital
parameters set this X-ray source apart from typical ABs, which have
shorter orbital periods and circular orbits. Indeed, the location of
this binary in the diagnostic diagrams of \citet[][Figures 5a and
  5b]{verbuntphinney95} indicate the orbit is too wide for the primary
to have circularised the orbit within the age of the cluster. However,
since the time scale for tidal synchronisation is shorter than for
circularisation, the rotation of one or both stars in this binary can
already be faster compared to single (sub-)giants or turnoff stars in
NGC\,188, which would explain the X-ray emission. The
pseudo-synchronisation period, defined in \citet{Hut:1981p99} to be
the spin period of the star if it corotates with the average orbital
speed around periastron, is 13.1 days in this binary. This is a bit
longer than the corotation period at periastron passage (10.6
days). CX\,13 may be an RS\,CVn-like system like the binary and X-ray
source S\,1242 in M\,67, which is an eccentric ($e=0.66$)
$P_{sp}=31.8$ day period binary on the subgiant branch.

WOCS\,4705 is the optical variable V\,11. The observed 0.4 mag drop in
brightness led \citet{kaluzny90} to postulate that V\,11 is a possible
eclipsing RS\,CVn binary. Interestingly, \citet{geller2009}, based on
their orbital ephemeris, found that this dip in brightness occurred at
the orbital phase where an eclipse would indeed be expected to
occur. Subsequent variability studies
\citep{mazurkaluzny1990,zhang2004} only observed small-amplitude
variations (0.03--0.08 mag). \citet{zhang2004} reported a possible
photometric period of 1.2 days. We note that folding the published
photometry for V\,11 on this period does not produce a smooth light
curve; on the other hand, this period does give a better match to the
$L_X$ versus $P_{rot}$ trend in Figure \ref{ch3_pvslx}. If real, the
origin of this period is unexplained. CX\,13 lies within the core
radius of the cluster, where the number of spurious matches between
X-ray sources and photometric variables is expected to be $\sim0.013$,
making the chance alignment of CX\,13 with a variable very unlikely.

\subsection{CX\,84}\label{ch3_cx84}

CX\,84 is a source located 3$\farcm$7 from the cluster centre that is
identified with the BSS WOCS\,5379. From radial velocities
\cite{geller2009} derived an eccentric ($e=0.24\pm0.03$) orbit with a
period of $P_{sp}=120.21$ days. It is one of the seven BSSs in
NGC\,188 for which FUV observations with the {\em Hubble Space
  Telescope} uncovered an excess flux compared to what is expected
from the BSS alone, showing that the companion in these systems (all
spectroscopic binaries) is a white dwarf
\citep{gosnell2014,gosnell2015}. From the temperature of the white
dwarf in WOCS\,5379 ($T\approx17,600$ K) it was derived that it formed
only $\sim77\pm7$ Myr ago. \cite{gosnell2014} sketched a possible
evolutionary scenario, in which the original secondary became a BSS
after mass accretion from the white-dwarf progenitor. However, there
are two aspects left unexplained by this evolutionary path. It fails
to explain the highly eccentric orbit, as mass transfer via Roche-lobe
overflow should circularise the orbit of the binary quite rapidly
(although under certain circumstances this is not necessarily the
case, see \cite{sepinsky2007,sepinsky2009}). Also, the nature of the
binary components does not account for the X-rays: the white dwarf is
not hot enough for thermal X-ray emission, and it is not obvious why
the BSS would be an X-ray source. Another cluster BSS within the {\em
  Chandra} FOV, WOCS\,4540, with an age ($70\pm7$ Myr;
\citealt{gosnell2014}) similar to WOCS\,5379, was not detected as an
X-ray source.

The key to solving this puzzle may lie in the optical variability that
was detected for WOCS\,5379. \citet{kafhon2003} found this source to
be a photometric variable (WV\,3) with a $V$ amplitude of 0.22 mag and
a period of $P_{ph}=0.18148$ days (if the light curve has one
maximum/minimum per period), or instead $P_{ph}=0.36296$ days if there
are two maxima/minima per period. What is the origin of this
variability?  According to \citet{geller2009}, WOCS\,5379 lies outside
the instability strip, ruling out pulsations as a likely
explanation. Alternatively, the variability may be the signature of a
short-period binary ``hiding'' in the system. The distribution of the
data points around photometric minimum when folded on the shorter
period \citep{kafhon2003} looks bimodal, which may be a sign that the
actual period is $\sim$0.36 days and that the two minima are
uneven. The presence of a short-period binary in the system would
provide possible explanations for the X-rays (coronal activity,
low-level mass transfer). On the other hand, it remains unclear what
the exact configuration of such a hierarchical multiple system would
be and whether any of the observed system components (the BSS and the
white dwarf) are part of the short-period binary. \citet{geller2009}
report that they were unable to derive a kinematic orbital solution
for a period of 0.18148 days, and also do not find a sign of rapid
rotation in the spectrum of WOCS\,5379.  This implies that the BSS,
which dominates the optical spectrum, is not part of the short-period
binary. The white dwarf could have a close companion, perhaps similar
to the pre-cataclysmic variables. But if mass transfer from the
white-dwarf progenitor to the BSS progenitor in the wide orbit ended
only $\sim77$ Myr ago (as inferred from the cooling age of the white
dwarf), then this close companion entered the system only recently (in
an encounter that perhaps also induced the eccentricity), complicating
this speculative formation scenario even more. An extensive
  exploration of conceivable evolutionary scenarios is beyond the
  scope of this paper. In light of our findings, which suggest that
  CX\,84/WOCS\,5379 is not a binary but a triple or higher-order
  multiple, we conclude that its actual formation may be more
  complicated than the scenario of mass transfer between two binary
  components as sketched by \cite{gosnell2014}. In star clusters with
  a binary frequency exceeding 10\% (such as in NGC\,188 where the
  frequency of main-sequence binaries with periods up to $10^4$ days
  is 29\%$\pm$3\%; \citealt{geller2012}), binary--binary encounters
  typically dominate over binary--single-star encounters
  \citep{sigurdsson1993,leigh2011}, producing, among others,
  hierarchical triples. \cite{leigh2013} have demonstrated the
  significance of dynamical encounters involving triple systems,
  especially in open clusters. In fact, as has been argued by
  \cite{leigh2011}, a considerable fraction of the blue stragglers in
  NGC\,188 could have a dynamical origin. Such dynamical encounters
  can become quite complex: in low-mass star clusters such as old open
  clusters or low-mass globular clusters, the encounter durations are
  comparable to the time scale for another encounter to happen. As a
  result, the probability of an ongoing encounter being interrupted by
  a subsequent encounter can be up to a few tens of percent
  \citep{gellerleigh2015}. The same is found when binaries of
  hydrogen-rich, non-degenerate stars undergoing stable mass transfer
  are considered: in low-mass clusters, the encounter time scale is
  comparable to the duration of the mass-transfer phase, resulting in
  a significant probability that a mass-transfer binary could be
  disrupted or affected by other nearby stars
  \citep{leigh2016apj}. The outcomes of such events, or sequence of
  events, are difficult to predict for individual cases, but the point
  is that the possibility of dynamical interactions playing a role in
  the formation of CX\,84/WOCS\,5379 should not be neglected.

We conclude that WOCS\,5379 is an interesting target for follow-up
studies to improve the current light-curve and period
measurements. Also, high--signal-to-noise high-resolution spectra
would enable a search for activity or accretion signatures. By
subtracting a high-quality template spectrum for the BSS, one could
look for any additional components contributing to the optical
spectrum. A similar technique was used to uncover a close binary in
the spectrum of the BSS S\,1082 in M\,67 \citep{vandenBerg:2001p1001},
for which a long orbital period was found from radial velocities
($\sim1189$ days; \citealt{sandquist2003}) while at the same time
eclipses on a 1-day period had been reported \citep{goranskij+92}.

 \subsection{Comparison With Other Old Star Clusters}

We compare the number of X-ray sources that we find in NGC\,188 to
those found in other old open clusters. To compare these results
uniformly, we select sources brighter than $L_X\approx1\times10^{30}$
erg s$^{-1}$ (0.3--7~keV) that are found inside $r_h$. For NGC\,188,
using a 2~keV MeKaL model to estimate the X-ray fluxes, we observe 55
such sources. Of these 55 X-ray sources, 9 are confirmed cluster
members, and another 3 have uncertain membership classification (the
remaining ones are likely AGNs). Among these, four are ABs (CX\,13,
CX\,22, CX\,38, CX\,40), two are CV/AGN candidates (CX\,14, CX\,28),
two are SSGs (CX\,26, CX\,49), one is a BSS (CX\,53), one an FK\,Com
(CX\,4), and two sources have uncertain classification (CX\,33,
CX\,48). We present these numbers in Table\,\ref{ch3_tab}, and compare
them with four other old open clusters. We find that the addition of
NGC\,188 to the sample confirms the trend that the number of X-ray
sources appears to scale with mass in open clusters when the three
clusters with the highest-quality membership information (M\,67,
NGC\,188 and NGC\,6791) are considered. This is expected for a
population that is (mostly) of primordial origin. The number of ABs
does not show an obvious scaling with mass; the small source samples
inhibit making any firm conclusions.

It has been pointed out that globular clusters and other old stellar
populations appear to be deficient in X-ray luminosity per unit mass,
when compared to old open clusters (e.g.~\citealt{verbunt2000},
\citealt{vandenBerg:2013p442}, \citealt{geea2015}). X-ray emissivities
of various old populations have been reported in the recent
literature. Due to the variety of methods and energy bands adopted, it
is not straightforward to compare the resulting values directly. The
X-ray emissivities in \cite{geea2015} were computed in a consistent
manner. To enable a comparison of our result for NGC\,188 with their
set of X-ray emissivities of dwarf ellipticals and globular clusters,
we have converted the X-ray emissivity of NGC\,188 to the 0.5--2 keV
and 2--8 keV bands for the assumption of a 2~keV MeKaL model. We find
that $L_X/M \approx 15.9 \times 10^{27}$ erg s$^{-1}$ $M_{\odot}^{-1}$
(0.5--2 keV), which is indeed higher than the X-ray emissivities in
Table 2 from Ge et al. However in the 2--8 keV band, the extrapolated
NGC\,188 emissivity is $L_X/M \approx 8.7 \times 10^{27}$ erg s$^{-1}$
$M_{\odot}^{-1}$, which is included in the range of X-ray emissivities
in \cite{geea2015}. We caution that our adopted spectral model to
compute X-ray fluxes is different than the one in their paper, which
is a thermal model with a log-normal temperature distribution around a
(fitted) peak temperature value.  A uniform reanalysis of the X-ray
data of old populations lies outside the scope of this paper, but we
refer the reader to the forthcoming paper by Heinke et al.\,that makes
an in-depth comparison of X-ray emissivities of globular clusters and
those of other old stellar populations, including old open clusters.

We illustrate the elevated X-ray emission of old open clusters with
concrete numbers. NGC\,188 has a mass of $2600\pm460$ $M_{\odot}$
\citep{geller2008} and has two members (the FK\,Com star CX\,4, and
the subgiant AB CX\,13) with $L_X\gtrsim5\times 10^{30}$ erg s$^{-1}$
(0.3--7 keV) within $r_h$. In M\,67, with a mass of
2100$^{+610}_{-550}$ \citep{geller2015} there are three members above
this luminosity limit: a SSG, a BSS and an AB. Finally, in NGC\,6791
(5000--7000 $M_{\odot}$; \citealt{platea11}), seven to eight members
are detected above this luminosity limit; they are a mix of CVs, SSGs,
ABs and an apparently single red giant. In contrast, the two sparse
globular clusters Terzan\,3 and NGC\,6535 with masses $\sim$12\,000
$M_{\odot}$ were also observed down to $\sim5\times 10^{30}$ erg
s$^{-1}$ (0.3--7 keV) with {\em Chandra}, and zero sources were
detected inside $r_h$ (A.\,Kong, private communication). Several
factors can contribute to the enhanced X-ray emissivity of old open
clusters, including the faster rate at which stars are lost from open
clusters compared to globular clusters as a result of their shorter
relaxation times (reducing their mass more severely), the higher rate
of dynamical destruction in globular clusters of certain types of
binaries that contribute to the X-ray source populations of open
clusters, differences in age, and differences in metallicity (see also
the discussion in Section 5.2 of \citealt{vatsvdb2017}). Further
studies of more clusters with a range of properties are needed to gain
more insight.

\section{Summary} \label{ch3_summary}

We present the results of a new X-ray study of the 7-Gyr old open
cluster NGC\,188. {\em Chandra} detected 84 sources above
$L_X\approx4\times10^{29}$ erg s$^{-1}$ (0.3--7 keV), of which 73 lie
within the half-mass radius. Of the thirteen cluster members that were
detected by {\em Chandra}, seven are new X-ray detections. The X-ray
source population of this cluster is a mix of ABs, SSGs, BSSs, and a
single FK\,Com star. While some of these X-ray source types are
frequently seen in other old open clusters, we also found a few
surprises. CX\,33 is an apparently single cluster star on the main
sequence, and we do not understand its X-ray emission. CX\,13 (V\,11)
is an AB that may be on its way to being circularised. The orbital
parameters of the BSS and white-dwarf binary WOCS\,5379 already were
puzzling given the combination of an eccentric orbit in a
post--mass-transfer system, but the detection of X-rays (CX\,84) makes
this system even more intriguing. It raises the question whether the
short-period variability that has been reported for this star is
coming from a close binary inside the system, which would also explain
the X-ray emission. Considering the overall X-ray population in
NGC\,188, we confirm our findings that the X-ray emissivity of old
open clusters is elevated compared to other old stellar populations.

\section*{Acknowledgments}

The authors would like to thank A. Kong for sharing the {\em Chandra}
results for Terzan\,3 and NGC\,6535 prior to publication.  S.V. would
like to acknowledge support from NOVA (Nederlandse Onderzoekschool
Voor Astronomie). This research has made use of data obtained from the
3XMM {\em XMM-Newton} serendipitous source catalogue compiled by the
10 institutes of the {\em XMM-Newton} Survey Science Centre selected
by ESA. This work is supported by {\em Chandra} grant GO3-14098X.
 
{\it Facilities:} CXO

\begin{landscape}
\begin{table}
\caption{Catalogue of {\em Chandra} sources in NGC\,188}
\label{ch3_tab1}
\begin{tabular}{lcccccccclccc} \hline \hline
(1) & (2) & (3) & (4) & (5) & (6) &(7) & (8) & (9) &(10) &  (11) & (12) & (13)\\
CX & CXOUJ & $\alpha$ (J2000.0) & $\delta$ (J2000.0) & Error & $\theta$ & C$_{t,net}$ & C$_{s,net}$ & $F_{X,u}$ & $E_{50}$ & Optical  & X  & GX     \\ 
& & & & & & & &  & &Match & &\\  
 & & ($^{\text{o}}$) & ($^{\text{o}}$) & ($''$) & ($'$) & & & (10$^{-15}$ erg cm$^{-2}$ s$^{-1}$) & (keV) & & &\\ \hline 
1 & 005124.1+851803 &  12.850489 &  85.300911 &  0.52 &  6.05 & 256$\pm$16 & 200$\pm$14 & 123 & 1.29$\pm$0.04 & + &26$^{a}$ &1\\ 
2 & 005027.5+852211 &  12.614557 &  85.369736 &  0.78 &  8.08 & 189$\pm$13.5 & 131$\pm$12 & 109 & 1.50$\pm$0.04 & + & 19&3\\ 
3 & 004244.2+851414 &  10.684148 &  85.237289 &  0.63 &  5.47 &  74$\pm$8.7 &  72$\pm$9 & 39.1 & 0.94$\pm$0.07 &  +& 30&2\\ 
4 & 004757.0+851456 &  11.987457 &  85.248873 &  0.35 &  1.16 &  63$\pm$8.0 &  56$\pm$8 & 27.9 & 1.17$\pm$0.07 &  +& 29& - \\ 
5 & 004745.1+852211 &  11.937813 &  85.369615 &  0.97 &  6.97 &  54$\pm$7 &  29$\pm$5 & 32.3 & 1.8$\pm$0.2 &  --& 20 &13\\ 
6 & 004127.6+851636 &  10.364926 &  85.276781 &  1.06 &  7.06 &  46$\pm$6.9 &  26$\pm$5 & 31.5 & 1.8$\pm$0.2 &  +& - &4\\ 
7 & 004301.9+851309 &  10.757843 &  85.219053 &  0.75 &  5.44 &  41$\pm$6.5 &  21$\pm$5 & 21.2 & 1.9$\pm$0.4 &  +& - &14\\ 
8 & 004448.3+850845 &  11.201416 &  85.145926 &  1.13 &  7.09 &  40$\pm$6 &  21$\pm$5 & 24.5 & 2.0$\pm$0.2 &  --& - &40\\ 
9 & 005330.0+852412 &  13.375125 &  85.403279 &  3.79 & 11.90 &  38$\pm$6.6 &  25$\pm$5 & 46.3 & 1.4$\pm$0.2 &  --& - &23\\ 
10 & 005112.8+851813 &  12.803453 &  85.303481 &  0.90 &  5.92 &  35$\pm$6 &  25$\pm$5 & 16.9 & 1.41$\pm$0.19 &  --& - & 1 \\ 
11 & 004458.7+851919 &  11.244769 &  85.321953 &  0.70 &  4.80 &  33$\pm$6 &  20$\pm$5 & 15.5 & 1.7$\pm$0.2 &  +& - &29\\ 
12 & 004611.4+851444 &  11.547483 &  85.245451 &  0.38 &  1.20 &  31$\pm$6 &  24$\pm$5 & 13.7 & 1.47$\pm$0.13 & + & - &16\\ 
13 & 004522.6+851238 &  11.344142 &  85.210584 &  0.53 &  3.36 &  30$\pm$6 &  23$\pm$5 & 16.9 & 1.30$\pm$0.13 &  +& - &18\\ 
14 & 005212.2+851058 &  13.050737 &  85.182854 &  1.57 &  7.73 &  30$\pm$6 &  16$\pm$4 & 14.7 & 1.9$\pm$0.4 &  +& - & - \\ 
15 & 004508.8+850859 &  11.286548 &  85.149710 &  1.21 &  6.72 &  29$\pm$6 &  22$\pm$5 & 14.7 & 1.50$\pm$0.14 & + & - & - \\ 
16 & 005247.9+850843 &  13.199496 &  85.145399 &  2.82 &  9.73 &  27$\pm$6 &   8$\pm$3 & 21.1 & 2.6$\pm$0.2 &  --& - & - \\ 
17 & 004607.5+851349 &  11.531248 &  85.230399 &  0.42 &  1.85 &  27$\pm$5 &  21$\pm$5 & 11.8 & 1.08$\pm$0.06 & + & - & - \\ 
18 & 004223.8+851515 &  10.599279 &  85.254186 &  1.05 &  5.78 &  23$\pm$5 &  13$\pm$4 & 13.2 & 1.8$\pm$0.6 &  --& - & - \\ 
19 & 004056.4+851822 &  10.235198 &  85.306026 &  2.04 &  8.16 &  23$\pm$5 &  13$\pm$4 & 19.6 & 2.0$\pm$0.3 &  --& - &46\\ 
20 & 004647.0+851437 &  11.695984 &  85.243501 &  0.38 &  0.73 &  23$\pm$5 &  19$\pm$4 & 10.0 & 1.31$\pm$0.14 & -- & - & - \\ 
21 & 004956.3+852113 &  12.484570 &  85.353716 &  1.49 &  6.93 &  22$\pm$5 &  11$\pm$3 & 12.9 & 2.1$\pm$0.4 & -- & 21& 17 \\ 
22 & 005002.4+852123 &  12.510198 &  85.356311 &  1.58 &  7.13 &  22$\pm$5 &  19$\pm$5 & 11.8 & 1.12$\pm$0.09 & + & 21&17\\ 
23 & 004747.4+851105 &  11.947578 &  85.184604 &  0.72 &  4.28 &  22$\pm$5 &   10$\pm$3 & 10.9 & 2.5$\pm$0.6 &  --& - & - \\ 
24 & 005229.8+852315 &  13.123994 &  85.387501 &  3.96 & 10.40 &  22$\pm$5 &  10$\pm$4 & 14.7 & 2.4$\pm$0.4 & +& - &57\\ 
25 & 004256.6+851821 &  10.735748 &  85.305800 &  1.15 &  5.94 &  21$\pm$5 &   10$\pm$3 & 10.7 & 2.2$\pm$0.5 & -- &25 &36\\ 
26 & 004240.0+851649 &  10.666620 &  85.280363 &  1.14 &  5.65 &  19$\pm$4 &  16$\pm$4 & 11.2 & 1.23$\pm$0.18 & + & - &28\\ 
27 & 004131.7+851814 &  10.381958 &  85.303937 &  1.99 &  7.45 &  18$\pm$5 &   8$\pm$3 & 10.8 & 2.4$\pm$0.5 & -- & - &42\\ 
28 & 004415.7+851811 &  11.065291 &  85.302996 &  0.86 &  4.52 &  17$\pm$4 &  15$\pm$4 & 79.9 & 1.13$\pm$0.13 & + & - &31 \\ 
29 & 005151.5+852319 &  12.964422 &  85.388607 &  4.44 &  9.97 &  16$\pm$5 &   9$\pm$3 & 10.1 & 1.8$\pm$0.3 &  + & - & - \\ 
30 & 004058.6+851859 &  10.244231 &  85.316497 &  2.95 &  8.37 &  15$\pm$4 &   9$\pm$3 & 10.8 & 1.96$\pm$0.18 & -- & - & 54\\ 
31 & 005107.4+851735 &  12.781013 &  85.293161 &  1.26 &  5.54 &  15$\pm$4 &   7$\pm$3 & 7.26 & 2.0$\pm$0.6 &  -- & - & - \\ 
32 & 004525.7+852145 &  11.357141 &  85.362564 &  1.91 &  6.79 &  14$\pm$4 &   7$\pm$3 & 8.60 & 2.0$\pm$0.9 &  -- & - & - \\ 
33 & 005201.6+851739 &  13.006512 &  85.294043 &  1.80 &  6.59 &  14$\pm$4 &   9$\pm$3 & 7.03 & 1.5$\pm$0.3 &  + & - & - \\ 
34 & 004540.2+851616 &  11.417386 &  85.270972 &  0.49 &  1.99 &  14$\pm$4 &   8$\pm$3 & 6.86 & 2.0$\pm$0.5 &  -- & -& - \\ 
35 & 005257.2+851160 &  13.238268 &  85.199931 &  2.96 &  8.05 &  13$\pm$4 &   9$\pm$3 & 7.21 & 1.7$\pm$0.3 &  -- & - & - \\ 
36 & 004334.4+851429 &  10.893236 &  85.241471 &  0.95 &  4.40 &  13$\pm$4 &   7$\pm$3 & 6.26 & 1.9$\pm$0.5 & -- & - & - \\ 
37 & 004752.3+852016 &  11.968118 &  85.337910 &  1.19 &  5.11 &  13$\pm$4 &   5$\pm$2 & 6.97 & 2.2$\pm$0.4 &  -- & -& - \\ 
38 & 004429.9+852056 &  11.124682 &  85.348783 &  1.89 &  6.48 &  12$\pm$4 &  12$\pm$4 & 6.26 & 1.19$\pm$0.11 & + & - & 20\\ 
39 & 004530.0+852329 &  11.374944 &  85.391295 &  3.56 &  8.44 &  12$\pm$4 &   5$\pm$3 & 8.83 & 3$\pm$1 &  -- & 17& - \\ 
40 & 005044.9+851139 &  12.687138 &  85.194069 &  1.63 &  5.86 &  11$\pm$4 &   9$\pm$3 & 7.38 & 0.9$\pm$0.5 & + & - & - \\  \hline
\end{tabular}
\end{table}
\end{landscape}

\begin{landscape}
\begin{table}
\contcaption{Catalogue of {\em Chandra} sources in NGC\,188}
\begin{tabular}{lcccccccclccc} \hline \hline
(1) & (2) & (3) & (4) & (5) & (6) &(7) & (8) & (9) &(10) &  (11) & (12) & (13) \\
CX & CXOUJ & $\alpha$ (J2000.0) & $\delta$ (J2000.0) & Error & $\theta$ & C$_{t,net}$ & C$_{s,net}$ & $F_{X,u}$ & $E_{50}$ & Optical  & X  & GX   \\ 
& & & & & & & &  & &Match & &\\  
 & & ($^{\text{o}}$) & ($^{\text{o}}$) & ($''$) & ($'$) & & & (10$^{-15}$ erg cm$^{-2}$ s$^{-1}$) & (keV) & &  &\\ \hline 
41 & 004551.2+850922 &  11.463303 &  85.156187 &  1.75 &  6.08 &  11$\pm$4 &   4$\pm$2 & 5.58 & 2.9$\pm$0.6 &  -- & - & - \\ 
42 & 004507.5+851525 &  11.281261 &  85.256819 &  0.58 &  2.40 &  11$\pm$3 &   9$\pm$3 & 4.95 & 1.5$\pm$0.3 &  -- & - &25 \\ 
43 & 005327.1+851516 &  13.362833 &  85.254349 &  3.41 &  7.94 &  11$\pm$4 &   7$\pm$3 & 6.36 & 1.7$\pm$0.2 &  -- & - & - \\ 
44 & 005032.7+850824 &  12.636064 &  85.139905 &  3.70 &  8.14 &  11$\pm$4 &   3$\pm$2 & 5.79 & 2.8$\pm$0.7 &  -- & - & - \\ 
45 & 004224.5+850807 &  10.602004 &  85.135169 &  5.46 &  9.23 &  10$\pm$4 &   5$\pm$3 & 6.43 & 1.9$\pm$0.7 &  + & - & - \\ 
46 & 005352.6+851527 &  13.469009 &  85.257432 &  4.28 &  8.46 &  10$\pm$4 &   3$\pm$2 & 6.45 & 2.5$\pm$0.6 &  -- & - & - \\ 
47 & 004400.1+851534 &  11.000257 &  85.259436 &  0.90 &  3.80 &   10$\pm$3 &   2$\pm$2 & 5.52 & 2.5$\pm$0.3 &  -- & - & - \\ 
48 & 004736.2+851105 &  11.900707 &  85.184716 &  1.04 &  4.23 &   10$\pm$3 &   2$\pm$2 & 4.38 & 4.3$\pm$0.7 &  + & - & - \\ 
49 & 004822.6+851555 &  12.094082 &  85.265166 &  0.54 &  1.76 &   9$\pm$3 &   9$\pm$3 & 4.34 & 1.06$\pm$0.14 &  + & - & 45\\ 
50 & 004508.0+851313 &  11.283363 &  85.220383 &  0.77 &  3.14 &   9$\pm$3 &   4$\pm$2 & 5.01 & 2.3$\pm$0.5 &  -- & - & - \\
51 & 005245.1+851214 &  13.187981 &  85.203906 &  3.86 &  7.72 &   9$\pm$3 &   8$\pm$3 & 4.73 & 0.99$\pm$0.07 &  + & - & - \\ 
52 & 005214.5+851628 &  13.060271 &  85.274553 &  2.65 &  6.54 &   8$\pm$3 &   6$\pm$3 & 4.01 & 1.5$\pm$0.6 &  -- & - & - \\ 
53 & 004711.7+851332 &  11.798637 &  85.225445 &  0.56 &  1.74 &   8$\pm$3 &   8$\pm$3 & 3.42 & 1.0$\pm$0.2 &  + & - & - \\ 
54 & 004511.5+851312 &  11.297759 &  85.219941 &  0.81 &  3.11 &   8$\pm$3 &   2$\pm$2 & 4.03 & 2.3$\pm$0.9 &  -- & - & - \\ 
55 & 004536.0+851422 &  11.399931 &  85.239559 &  0.60 &  2.02 &   8$\pm$3 &   4$\pm$2 & 3.50 & 2$\pm$1 &  -- & - & - \\ 
56$\dagger$ & 004133.8+851231 &  10.390686 &  85.208479 &  3.65 &  7.38 &   8$\pm$3 &   6$\pm$3 & 5.41 & 1.2$\pm$0.8 &  + & 32& 47 \\ 
57 & 004340.9+851402 &  10.920561 &  85.233950 &  1.24 &  4.37 &   8$\pm$3 &   6$\pm$3 & 3.80 & 1.3$\pm$0.6 &  -- & - & - \\ 
58 & 005024.3+851407 &  12.601358 &  85.235264 &  1.23 &  4.32 &   8$\pm$3 &   3$\pm$2 & 4.20 & 3.0$\pm$0.7 &  -- & - &49 \\ 
59 & 005202.4+851429 &  13.009953 &  85.241512 &  2.62 &  6.24 &   7$\pm$3 &   5$\pm$2 & 3.48 & 1.4$\pm$0.8 &  -- & - & - \\ 
60 & 004612.2+851402 &  11.550944 &  85.233869 &  0.57 &  1.62 &   7$\pm$3 &   7$\pm$3 & 3.08 & 1.11$\pm$0.16 & + & - & - \\ 
61 & 005044.4+851739 &  12.684838 &  85.294036 &  1.85 &  5.14 &   7$\pm$3 &   2$\pm$2 & 3.31 & 2.3$\pm$0.5 &  -- & - & - \\ 
62 & 004506.6+851029 &  11.277525 &  85.174860 &  2.05 &  5.35 &   6$\pm$3 &   3$\pm$2 & 3.05 & 3.2$\pm$1.0 &  -- & - & - \\ 
63 & 005200.6+851443 &  13.002680 &  85.245233 &  2.81 &  6.18 &   6$\pm$3 &   3$\pm$2 & 3.11 & 2.1$\pm$0.9 &  -- & - & - \\ 
64 & 004714.2+851408 &  11.809200 &  85.235518 &  0.56 &  1.15 &   6$\pm$3 &   3$\pm$2 & 2.60 & 2.6$\pm$2.0 &  -- & - & - \\ 
65 & 004704.3+851502 &  11.767804 &  85.250416 &  0.51 &  0.24 &   6$\pm$3 &   6$\pm$3 & 2.64 & 1.14$\pm$0.13 & + & - & - \\ 
66$\dagger$ & 004240.6+851934 &  10.669318 &  85.326108 &  3.90 &  6.90 &   6$\pm$3 &   2$\pm$2 & 3.31 & 3.5$\pm$0.8 &  -- & - & - \\ 
67 & 004654.0+851437 &  11.724808 &  85.243536 &  0.53 &  0.68 &   6$\pm$3 &   5$\pm$2 & 2.62 & 1.4$\pm$0.4 &  + & - & - \\ 
68 & 004550.0+851257 &  11.458336 &  85.215886 &  0.84 &  2.76 &   6$\pm$3 &   4$\pm$2 & 2.64 & 2$\pm$1 &  -- & - & - \\ 
69$\dagger$ & 005028.8+852004 &  12.620079 &  85.334540 &  3.47 &  6.40 &   6$\pm$2 &   2$\pm$2 & 3.16 & 2.9$\pm$0.9 & -- & - & - \\ 
70$\dagger$ & 005011.2+851552 &  12.546482 &  85.264322 &  1.42 &  3.93 &   5$\pm$2 &  0. & 4.46 & 3.7$\pm$0.7 &  -- & - & - \\ 
71$\dagger$ & 005049.3+851611 &  12.705503 &  85.269756 &  2.02 &  4.76 &   5$\pm$2 &   5$\pm$2 & 2.37 & 1.19$\pm$0.19 & + & - & - \\ 
72$\dagger$ & 004353.5+851660 &  10.972969 &  85.283284 &  1.67 &  4.28 &   5$\pm$2 &   3$\pm$2 & 2.52 & 2.0$\pm$0.6 &  -- & - & - \\ 
73 & 004920.7+852017 &  12.336173 &  85.337920 &  3.07 &  5.75 &   5$\pm$2 &   2$\pm$2 & 2.23 & 3.6$\pm$2.0 &  -- & - & - \\ 
74 & 005138.7+851315 &  12.911108 &  85.220731 &  3.48 &  6.06 &   5$\pm$2 &   3$\pm$2 & 2.80 & 1.8$\pm$0.5 &  -- & - & - \\ 
75 & 004955.8+851910 &  12.482488 &  85.319465 &  2.53 &  5.27 &   5$\pm$2 &   5$\pm$2 & 2.34 & 1.3$\pm$0.3 &  + & - & - \\ 
76$\dagger$ & 005424.4+850928 &  13.601545 &  85.157819 & 19.80 & 10.90 &   5$\pm$3 &   6$\pm$3 & 3.24 & 1.6$\pm$0.3 &  + & - & - \\ 
77$\dagger$ & 005348.8+851046 &  13.453139 &  85.179525 & 13.30 &  9.57 &   5$\pm$3 &   7$\pm$3 & 2.77 & 0.86$\pm$0.12 & + & - & - \\ 
78 & 004413.9+851633 &  11.057838 &  85.275946 &  1.54 &  3.73 &   4$\pm$2 &   4$\pm$2 & 1.83 & 1.11$\pm$0.05 &  + & - & - \\ 
79$\dagger$ & 004454.0+851841 &  11.225075 &  85.311377 &  2.02 &  4.33 &   4$\pm$2 &   2$\pm$2 & 1.74 & 1.5$\pm$0.7 &  -- & - & - \\ 
80$\dagger$ & 004533.1+851343 &  11.387968 &  85.228607 &  1.11 &  2.43 &   3$\pm$2 &   1$\pm$1 & 1.35 & 5$\pm$1 &  + & - & - \\  \hline
\end{tabular}
\end{table}
\end{landscape}

\begin{landscape}
\begin{table}
\contcaption{Catalogue of {\em Chandra} sources in NGC\,188}
\begin{tabular}{lcccccccclccc} \hline \hline
(1) & (2) & (3) & (4) & (5) & (6) &(7) & (8) & (9) &(10) &  (11) & (12) & (13) \\
CX & CXOUJ & $\alpha$ (J2000.0) & $\delta$ (J2000.0) & Error & $\theta$ & C$_{t,net}$ & C$_{s,net}$ & $F_{X,u}$ & $E_{50}$ & Optical  & X  & GX   \\ 
& & & & & & & &  & &Match & &\\  
 & & ($^{\text{o}}$) & ($^{\text{o}}$) & ($''$) & ($'$) & & & (10$^{-15}$ erg cm$^{-2}$ s$^{-1}$) & (keV) & &  &\\ \hline 
81 & 004756.3+851610 &  11.984630 &  85.269310 &  0.84 &  1.42 &   3$\pm$2 &   1$\pm$1 & 1.64 & 3$\pm$1 &  -- & - & - \\ 
82$\dagger$ & 004843.3+851203 &  12.180331 &  85.200703 &  2.00 &  3.83 &   3$\pm$2 &   0. & 1.33 & 6$\pm$1 &  -- & - & - \\ 
83$\dagger$ & 004916.2+851346 &  12.317489 &  85.229490 &  1.49 &  3.13 &   3$\pm$2 &   3$\pm$2 & 1.30 & 1.3$\pm$0.3 & -- & - & - \\ 
84 & 005011.2+851439 &  12.546689 &  85.244166 &  2.09 &  3.94 &   3$\pm$2 &   3$\pm$2 & 1.29 & 1.17$\pm$0.13 &  + & - & - \\ \hline
\multicolumn{13}{p {18cm}}{} \\   
\multicolumn{13}{p {20cm}}{Col.\,(1): X-ray catalogue sequence number, sorted by net X-ray counts (0.3--7~keV). Sources that were detected by {\tt wavdetect} using a {\tt sigthresh} of $10^{-6}$ but not with a {\tt sigthresh} of $10^{-7}$ have been flagged with $\dagger$ }\\
\multicolumn{13}{p {18cm}}{Col.\,(2): IAU designated source name }                   \\
\multicolumn{13}{p {18cm}}{Cols.\,(3) and (4): Right ascension and declination (in decimal degrees) for epoch J2000.0} \\
\multicolumn{13}{p {18cm}}{Col.\,(5): 95\% confidence radius on {\tt wavdetect} X-ray source position in arcseconds}  \\
\multicolumn{13}{p {18cm}}{Col.\,(6): Angular offset from the cluster centre ($\alpha_{2000} =
0^{h}47^{m}12.5^{s}$, $\delta_{2000} = +85^{\circ}14'\,49''$;
\citealt{Chumak:2010p1841}) in arcminutes}\\
\multicolumn{13}{p {18cm}}{Col.\,(7): Net counts extracted in the broad energy band (0.3--7~keV) with 1-$\sigma$ errors}\\
\multicolumn{13}{p {18cm}}{Col.\,(8): Net counts extracted in the soft energy band (0.3--2~keV) with 1-$\sigma$ errors}\\
\multicolumn{13}{p {18cm}}{Col.\,(9): Unabsorbed X-ray flux in the 0.3--7~keV energy band for a 2~keV MeKaL model and neutral hydrogen column of 5$\times$10$^{20}$ cm$^{-2}$}\\
\multicolumn{13}{p {18cm}}{Col.\,(10): Median energy $E_{50}$ in keV and 1-$\sigma$ errors on them.}\\
\multicolumn{13}{p {18cm}}{Col.\,(11) : Information about presence (+) or absence (--) of optical counterpart (details in Table 2)}\\
\multicolumn{13}{p {18cm}}{Col.\,(12) : ID of $ROSAT$ counterpart \citep{Belloni:1998p1070}; $^{a}$ -- outside the 95\% error circle but within 4-$\sigma$ match radius}\\ 
\multicolumn{13}{p {18cm}}{Col.\,(13) : ID of {\em XMM-Newton} counterpart \citep{Gondoin:2005p1051}}\\
\end{tabular}
\end{table}
\end{landscape}

%
% Table 2
%
\begin{landscape}
\begin{table}
\caption{Primary {\em Chandra} catalogue: Optical Counterpart Properties}
\label{ch3_tab2}
\begin{tabular}{rlllllllllccccl} \hline \hline              
(1)  & (2)                &   (3)        & (4)        & (5)       &  (6)     &(7)                        & (8)            & (9)               &(10)               & (11)                          & (12)                                 & (13)                                       &  (14)      \\
CX & OID              & Dox        & $V$      & $B-V$  & Var     & $P_{Ph}$            & $P_{Sp}$ & MP$_{PM}$ & MP$_{RV}$ & RV Variable                         & $L_{X,u}$                & \textit{$\log(F_X/F_V)_{u}$} &Class \\
      &                     &  &             &             &           &                   &     &                     &                     &            Type         & &                                               &              &          \\
            &                     & ($''$) &             &             &           & (days)                  & (days)      &                     &                     &                                 &(10$^{30}$ erg s$^{-1}$) &                                               &              \\ \hline
\multicolumn{13}{p {17cm}}{Members } \\ \hline
4   & 5027            &  0.0521   & 11.939 &  1.202  &   $\ldots$  &         $\ldots$      &    $\ldots$  &        96         &         98        & SM                          &     9.08&  --3.44& FK\,Com                                \\ 
13  & 4705            &  0.2194   & 13.964 &  0.962  & V11    &    1.2433$^z$       & 35.1780   &        98         &         98        & BM                          &     5.51&  --2.85& AB       \\ 
22  & 5258            &  0.3436   & 15.569 &  0.753  & V21     & 1.3836                &   $\ldots$    &          92        &    $\ldots$   & U                            &     3.84&  --2.36& AB                                       \\ 
%      &                     &                &             &             &             & 1.17254$^z$      &                 &                     &                     &                                 &                                        &                                                            & \\ 
26  & 4289            &  0.5519   & 15.326 &  0.984  & $\ldots$   &  $\ldots$       & 11.4877   &         98        &         98        & BM                          &     3.64&  --2.48& SSG                                       \\ 
33  & 5639            &  0.5709   & 16.038 &  0.738  & $\ldots$  &      $\ldots$     & $\ldots$  &         90        &         98        & SM                          &     2.29&  --2.40& Unc                           \\ 
38  & 4508            &  0.8726   & 16.121 &  0.751  & WV28 &   0.97157$^k$      &  $\ldots$   &           98      &  $\ldots$     & BU                           &     2.04&  --2.42& AB                                       \\ 
40  & 5459            &  0.6071   & 17.518 &  0.998  & V10    &   3.1596               &  $\ldots$  &         94        &   $\ldots$  &    $\ldots$      &     2.41&  --1.79& AB                                       \\ 
49  & 4989            &  0.2781   & 16.104 &  0.970  & V05    &   0.5860               &   $\ldots$   &         95        &   $\ldots$    &     $\ldots$          &     1.41&  --2.58& SSG                      \\ 
%      &                     &                &             &             &           &0.5859835$^z$     &                 &                     &                     &                                &                                        &                                                &              &\\ 
53  & 5078            &  0.0727   & 14.465 &  0.601  & V34    &   4.5295               & 4.78303   &         98        &           97      & BM                          &     1.11&  --3.34& BSS                                       \\ 
65  & 5024            &  0.1649   & 17.071 &  0.957  & V33    &    3.2064              &  $\ldots$  &         93        &    $\ldots$   &       $\ldots$    &     0.860&  --2.41& AB                                       \\ 
71  & 5337            &  0.8961   & 15.759 &  0.848  & V04    &   0.3425               &   $\ldots$   &         96        &     $\ldots$   & U                            &     0.772&  --2.98& AB-W\,UMa                              \\ 
 %     &                     &                &             &             &            &0.342459697$^z$&                 &                     &                     &                                &                                        &                                                &               &\\ 
78  & 4603            &  0.5702   & 17.753 &  1.140  & V37     &  7.0235               &  $\ldots$  &         91        &  $\ldots$      &      $\ldots$       &     0.595&  --2.30&AB                                       \\ 
84  & 5379            &  1.0484   & 15.372 &  0.570  & WV3    &   0.18148$^k$    & 120.2100  &        98         &          98      & BM                          &     0.421&  --3.40& BSS                                       \\ \hline
\multicolumn{13}{p {17cm}}{Uncertain Members } \\ \hline
14  & 5760            &  1.4338   & 21.462 &  0.276  &$\ldots$&       $\ldots$    &   $\ldots$    &       $\ldots$   &   $\ldots$    &     $\ldots$      &     4.78&   0.09& CV/AGN candidate      \\ 
24  & 5569            &  1.5544   & 21.865 & --0.287 & WV2   &   0.15059$^k$      &    $\ldots$   &    $\ldots$      &   $\ldots$     &     $\ldots$            &     4.79&   0.25& CV candidate                    \\ 
28  & 4566            &  0.5028   & 21.904 &  0.099  &$\ldots$&       $\ldots$    & $\ldots$   &   $\ldots$       &  $\ldots$     &      $\ldots$              &     2.60&   0.00& CV/AGN candidate                         \\ 
29  & 5565            &  1.2372   & 20.979 &  0.381  &$\ldots$&     $\ldots$       &$\ldots$ &     $\ldots$   &    $\ldots$     &      $\ldots$              &     3.29&  --0.27& CV/AGN candidate                    \\ 
45  & 4428            &  1.4499   & 21.460 &  0.862  &$\ldots$&       $\ldots$        &     $\ldots$     &$\ldots$&$\ldots$&$\ldots$&     2.09                            &  --0.27& CV/AGN candidate                  \\ 
48  & SMV-8056   &  0.5807   & 20.721 &  1.614  &$\ldots$&      $\ldots$         &    $\ldots$    &$\ldots$&$\ldots$&$\ldots$&     1.43                            &  --0.73&       Unc                  \\ 
76  & 7237            &  1.6723   & 21.737 &  0.289  &$\ldots$&         $\ldots$           &    $\ldots$    &$\ldots$&$\ldots$&$\ldots$&     1.06                            &  --0.46& CV/AGN candidate                              \\ \hline
\multicolumn{13}{p {17cm}}{Non-Members } \\ \hline
 1   & 5629            &  0.0414   & 13.293 &  0.800  & V08    & 5.2096$^z$         & 4.08675   &            0        &            0        & BN&$\ldots$&  --2.26&Foreground RSCVn                         \\ 
 2   & 5250            &  0.3049   & 17.840 &  0.346  &    $\ldots$       &$\ldots$&$\ldots$&            0        &$\ldots$&$\ldots$&$\ldots$&  --0.49& AGN candidate                    \\ 
 3   & 877-000366 &  0.9800   &  9.525 &  0.678  &$\ldots$&$\ldots$&$\ldots$&$\ldots$&$\ldots$&$\ldots$&$\ldots$&  --4.26                                               & Foreground star                              \\ 
 6   & 4035            &  0.2945   & 17.776 &  0.421  &$\ldots$&$\ldots$&$\ldots$&            0        &$\ldots$&$\ldots$&$\ldots$&  --1.05& AGN candidate                   \\ 
 7   & 4354            &  0.2478   & 20.228 &  0.407  &$\ldots$&$\ldots$&$\ldots$&            0       &$\ldots$&$\ldots$&$\ldots$&  --0.25                                            & AGN candidate                  \\ 
11  & 4541            &  0.2566   & 21.340 &  0.020  &$\ldots$&$\ldots$&$\ldots$&           0        &$\ldots$&$\ldots$&$\ldots$&   0.06                                               & AGN candidate                \\ 
12  & 4645            &  0.1641   & 21.001 &  0.233  &$\ldots$&$\ldots$&$\ldots$&           0        &$\ldots$&$\ldots$&$\ldots$&  --0.13& AGN candidate\\ 
15  & 4764            &  0.4219   & 19.903 &  0.385  &$\ldots$&$\ldots$&$\ldots$&            0        &$\ldots$&$\ldots$&$\ldots$&  --0.53                              & AGN candidate\\ 
17  & 4679            &  0.3701   & 19.182 &  1.657  & WV29 &   0.97403$^k$     &$\ldots$&            0       &$\ldots$&$\ldots$&$\ldots$&  --0.92                                  & NM                              \\ 
51  & 5733            &  0.9698   & 15.985 &  0.765  & V29    &   1.1513               & 8.70400   &           98       &0     &BN&$\ldots$&  --2.59& AB            \\ \hline
\end{tabular}
\end{table}
\end{landscape}

\begin{landscape}
\begin{table}
\contcaption{Primary {\em Chandra} catalogue: Optical Counterpart Properties}
\begin{tabular}{rlllllllllccccl} \hline \hline              
(1)  & (2)                &   (3)        & (4)        & (5)       &  (6)     &(7)                        & (8)            & (9)               &(10)               & (11)                          & (12)                                 & (13)                                       &  (14)      \\
CX & OID              & Dox        & $V$      & $B-V$  & Var     & $P_{Ph}$            & $P_{Sp}$ & MP$_{PM}$ & MP$_{RV}$ & RV Variable                         & $L_{X,u}$                & \textit{$\log(F_X/F_V)_{u}$} &Class \\
      &                     &  &             &             &           &                   &     &                     &                     &            Type         & &                                               &              &          \\
            &                     & ($''$) &             &             &           & (days)                  & (days)      &                     &                     &                                 &(10$^{30}$ erg s$^{-1}$) &                                               &              \\ \hline
\multicolumn{13}{p {17cm}}{Non-members: {\em continued} } \\ \hline
56  & 4082            &  0.2260   & 19.336 &  1.665  &$\ldots$&$\ldots$&$\ldots$&             0       &$\ldots$&$\ldots$&$\ldots$&  --1.20                                            & NM                              \\ 
60  & 5912            &  0.2783   & 16.386 &  0.831  &$\ldots$&$\ldots$&$\ldots$& 0       &$\ldots$&$\ldots$&$\ldots$&  --2.62                   & NM          \\ 
67  & 5934            &  0.4161   & 14.342 &  0.456  & V30\color{white}c\color{black}    &   1.2996\color{white}cc\color{black}                &$\ldots$&            0        &$\ldots$& BU&$\ldots$&  --3.51& NM          \\ 
75  & 5292            &  1.5376   & 13.075 &  0.545  &$\ldots$&$\ldots$&$\ldots$&             0       &            0        & SN&$\ldots$&  --4.06& NM                               \\ 
76  & 6203\color{white}xxxxxx\color{black}            & 18.5846  & 15.256 &  0.524  &$\ldots$&$\ldots$&$\ldots$\color{white}ccccc\color{black}&             0       &$\ldots$&$\ldots$&$\ldots$&  --3.05                                           & NM                               \\ 
77  & 6193            &  1.0664   & 14.560 &  0.846  &$\ldots$&$\ldots$&$\ldots$&            0\color{white}c\color{black}        &$\ldots$& U                              &$\ldots$&  --3.40                            & NM                               \\ 
80  & 4681            &  0.8204   & 18.330 &  1.206  &$\ldots$&$\ldots$&$\ldots$&            0        &$\ldots$&$\ldots$&$\ldots$&  --2.20                                              &  Unc                   \\ \hline
\multicolumn{14}{p {17cm}}{}\\       
\multicolumn{14}{p {22.5cm}}{Col.\,(1): X-ray catalogue sequence number }\\
\multicolumn{14}{p {22.5cm}}{Col.\,(2): Optical source ID. For CX\,48, the optical ID is from the Stetson catalogue \cite{stetson2004} as this source was not present in the Platais catalogue \cite{platais2003}, and for CX\,3, the optical ID is from the USNO CCD Astrograph Catalog ($UCAC$) \cite{Zacharias:2013p938}}\\
\multicolumn{14}{p {22.5cm}}{Col.\,(3): Distance between the X-ray source and its optical counterpart in arcsec. }\\
\multicolumn{14}{p {22.5cm}}{Col.\,(4): $V$ magnitude}\\
\multicolumn{14}{p {22.5cm}}{Col.\,(5): $B-V$ colour.}\\
\multicolumn{14}{p {22.5cm}}{Col.\,(6): Short-period binary counterpart ID \citep{Hoffmeister64,kaluznyshara1987,kaluzny90,zhang2004,kafhon2003}.}\\
\multicolumn{14}{p {22.5cm}}{Col.\,(7): Photometric period (in days) of the short-period binary counterpart. Values from \citealt{mochejska2008} unless specified -- $^{k}$\citealt{kafhon2003}; $^{z}$\citealt{zhang2004})}\\
\multicolumn{14}{p {22.5cm}}{Col.\,(8): Spectroscopic period (in days) of the short-period binary counterpart \citep{geller2009} }\\
\multicolumn{14}{p {22.5cm}}{Col.\,(9): Membership probability determined from proper-motion studies \citep{platais2003} }\\
\multicolumn{14}{p {22.5cm}}{Col.\,(10): Membership probability determined from radial velocity studies \citep{geller2008}}\\
\multicolumn{14}{p {22.5cm}}{Col.\,(11): Radial-velocity (RV) variable type \citep{geller2008}; SM - single member, BM - binary member, U - unknown class, BU - binary with unknown membership, SN - single non-member, BN - binary non-member}\\
\multicolumn{14}{p {22.5cm}}{Col.\,(12): Unabsorbed X-ray luminosity (0.3--7~keV), assuming the source lies at the distance of the cluster (1650 pc)}\\
\multicolumn{14}{p {22.5cm}}{Col.\,(13): Unabsorbed X-ray (0.3--7~keV) to optical ($V$ band) flux ratio using a 2~keV MeKaL model and neutral hydrogen column of $5\times10^{20}$ cm$^{-2}$}\\
\multicolumn{14}{p {22.5cm}}{Col.\,(14): Object classification : CV - Cataclysmic variable ; AB - Active binary ; SSG - Sub-subgiant ; BSS - Blue straggler star ; Unc - Uncertain classification; NM - Single or binary foreground or background (non-member) star; AGN - active galactic nucleus}\\
\end{tabular}
\end{table}
\end{landscape}

%
% Table 3
%
\begin{table*}
 \caption{Cross-reference table between our {\em Chandra} catalogue and the 3XMM-DR6 catalogue}
  \label{ch3_tab3}
 \begin{center}
 \begin{tabular}{cccc} \hline \hline              
CX  &  SRCID  & Flux & GX \\ 
 & & (10$^{-15}$ erg cm$^{-2}$ s$^{-1}$) & \\ \hline
3  & 201006401010002 & 91$\pm$2 & 2  \\
4  & 201006401010022 & 21$\pm$1 & - \\
5  & 201006401010014 & 16.3$\pm$0.7 & 13 \\
6  & 201006401010004 & 29.0$\pm$0.9 & 4  \\
7  & 201006401010020 & 24$\pm$1 & 14 \\
8  & 201006401010048 & 15$\pm$1 & 40 \\
9  & 201006401010038 & 17$\pm$1 & 23 \\
11 & 201006401010031 & 8.1$\pm$0.5 & 29 \\
12 & 201006401010019 & 15.3$\pm$0.8 & 16 \\
13 & 201006401010027 & 15.2$\pm$0.8 & 18 \\
15 & 201006401010074 & 11$\pm$1 & -  \\
17 & 201006401010047 & 6.9$\pm$0.6 & -  \\
18 & 201006401010075 & 4.5$\pm$0.4 & -  \\
19 & 201006401010041 & 6.7$\pm$0.5 & 46 \\
24 & 201006401010063 & 9.3$\pm$0.8 & 57 \\
25 & 201006401010045 & 8.3$\pm$0.5 & 36 \\
26 & 201006401010034 & 11.9$\pm$0.6 & 28 \\
27 & 201006401010033 & 7.3$\pm$0.5 & 42 \\
29 & 201006401010101 & 7.9$\pm$0.7 & -  \\
30 & 201006401010053 & 6.4$\pm$0.5 & 54 \\
34 & 201006401010162 & 2.0$\pm$0.3 & -  \\
35 & 201006402010183 & 1.6$\pm$0.9 & -  \\
36 & 201006401010066 & 3.8$\pm$0.5 & -  \\
37 & 201006401010093 & 3.0$\pm$0.4 & -  \\
38 & 201006401010026 & 13.0$\pm$0.6 & 20 \\
39 & 201006401010084 & 5.1$\pm$0.5 & -  \\
45 & 201006401010133 & 4$\pm$1 & -  \\
49 & 201006401010055 & 7.1$\pm$0.6 & 45 \\
51 & 201006402010136 & 10$\pm$2 & -  \\
55 & 201006402010140 & 2.2$\pm$0.5 & -  \\
56 & 201006401010046 & 15.0$\pm$0.8 & 47 \\
58 & 201006401010072 & 6.4$\pm$0.7 & 49 \\
59 & 201006401010128 & 5.0$\pm$0.8 & -  \\
64 & 201006401010119 & 3.3$\pm$0.5 & -  \\
72 & 201006401010122 & 2.6$\pm$0.3 & -  \\
74 & 201006401010148 & 2.7$\pm$0.6 & -  \\
76 & 201006401010123 & 9$\pm$2 & -  \\
76 & 201006402010151 & 7$\pm$2 & -  \\
77 & 201006401010080 & 13$\pm$2 & -  \\
78 & 201006402010175 & 1.4$\pm$0.4 & -  \\
79 & 201006401010155 & 2.9$\pm$0.5 & - \\ \hline
  \multicolumn{4}{l}{}                                             \\       
\multicolumn{4}{p {8.5cm}}{Col.\,(1): {\em Chandra} ID}         \\
\multicolumn{4}{p {8.5cm}}{Col.\,(2): {\em XMM-Newton} serendipitous source catalogue ID}         \\
\multicolumn{4}{p {8.5cm}}{Col.\,(3): Flux obtained from {\em XMM-Newton} (0.5 -- 4.5 keV)}         \\
\multicolumn{4}{p {8.5cm}}{using a power law model with $\Gamma=1.7$ and $N_{H}=3\times$10$^{20}$ cm$^{-2}$}         \\
\multicolumn{4}{p {8.5cm}}{Col.\,(4): X-ray source ID from \citet{Gondoin:2005p1051} }        \\
      \end{tabular}
      \end{center}
\end{table*}

%
% Table 4
%
 \begin{table*}
 \caption{Comparison of X-ray Sources in Old Open Clusters with $L_X\geq10^{30}$ erg s$^{-1}$ (0.3--7~keV) inside $r_h$}
 \label{ch3_tab}
\begin{center}
 \begin{tabular}{lccccccc} \hline \hline              
Cluster         &  Age     & Mass                   & $N_X$             & $N_{X,CV}$   & $N_{X,SSG}$  & $N_{X,AB}$   & log(2$L_{30}$/Mass) \\ 
                & (Gyr)    & ($M_{\odot}$)        &                   &             &             &             & \\
\hline
 M 67$^1$       & 4        & 2100$_{-550}^{+610}$  & 12                & 0           & 1           & 7 -- 8      & 28.6  \\
 NGC\,188$^2$       & 7        & 2300$\pm$460        & 9 -- 12            & $\leq2$     & 2           & 4           & 28.4 -- 28.5 \\
 NGC 6819$^3$   & 2 -- 2.4 & 2600                & 3 -- 8             & $\lesssim$1 & $\lesssim$1 & $\lesssim$4 & 28.8 -- 29.3 \\ 
 NGC 6791$^4$   & 8        & 5000 -- 7000        & 15 -- 19           & 3 -- 4       & 3          & 7 -- 11     & 28.6 -- 28.8\\
 Cr\,261$^{5,a}$ & 7        & 5800 -- 7200        & $\lesssim$26$\pm$8 & $\lesssim$4 & $\lesssim$2 & 2 -- 23     & $\lesssim$28.6 -- 28.7 \\
 \hline %check how you wrote it in cr261 paper
 
  \multicolumn{8}{l}{}                                             \\       
\multicolumn{8}{p {13cm}}{Col.\ (1): Cluster name}         \\
\multicolumn{8}{p {13cm}}{Col.\ (2): Cluster age in Gyr}         \\
\multicolumn{8}{p {13cm}}{Col.\ (3): Mass of the cluster in $M_{\odot}$}         \\
\multicolumn{8}{p {13cm}}{Col.\ (4): Number of X-ray sources within $r\leq r_h$ and $L_X \geq 1\times10^{30}$ erg s$^{-1}$}         \\
\multicolumn{8}{p {13cm}}{Col.\ (5): Number of CVs within aforementioned radius and luminosity limit}         \\
\multicolumn{8}{p {13cm}}{Col.\ (6): Number of SSGs within aforementioned radius and luminosity limit}         \\
\multicolumn{8}{p {13cm}}{Col.\ (7): Number of ABs within aforementioned radius and luminosity limit}         \\
\multicolumn{8}{p {13cm}}{Col.\ (7): Ratio of the total X-ray luminosity of sources inside $r_h$ brighter than $1\times10^{30}$ erg s$^{-1}$ ($L_{30}$), and cluster mass. The multiplicative factor 2 is included to scale the mass estimate to the half-mass radius.}         \\
\multicolumn{8}{p {13cm}}{References -- $^1$\cite{vandenBerg:2004p1040,geller2015}, $^2$this work and \cite{Chumak:2010p1841}, $^3$\cite{Gosnell:2012p685, platea13}, $^4$\cite{vandenBerg:2013p442,platea11}, $^5$\cite{vatsvdb2017}}\\
\multicolumn{8}{p {13cm}}{$^{a}$Membership information is unavailable for Cr\,261, leading to
large uncertainties.} \\

\end{tabular}

  \end{center}

 \end{table*}

%NOTES:
%
\clearpage
%%%%%%%%%%%%%%%%%%%%%%%%%%%%%%%%%%%%%%%%%%%%%%%%%%



%%\bibliographystyle{mnras}
%%\bibliography{/Users/smritivats/Research/Projects/biball.bib} % if your bibtex file is called example.bib

%**

%**

%\begin{thebibliography}{2}
%\color{green}\bibitem[\protect\citeauthoryear{Author}{2012}]{Author2012}
%Others S., 2012, Journal of Interesting Stuff, 17, 198 \color{black}
%\end{thebibliography}

%%%%%%%%%%%%%%%%%%%%%%%%%%%%%%%%%%%%%%%%%%%%%%%%%%

%%%%%%%%%%%%%%%%% APPENDICES %%%%%%%%%%%%%%%%%%%%%

%\appendix



%%%%%%%%%%%%%%%%%%%%%%%%%%%%%%%%%%%%%%%%%%%%%%%%%%

% Don't change these lines
\bsp	% typesetting comment
\label{lastpage}
\end{document}